\documentclass[reprint,superscriptaddress,amsmath,amssymb,aps,nofootinbib]{revtex4-2}
\usepackage{dsfont}
\usepackage{graphicx}
\usepackage[caption=false]{subfig}

\allowdisplaybreaks

\usepackage{dcolumn}
\usepackage{bm}
\usepackage{color}
\usepackage{xcolor}
\usepackage{epsfig}
\usepackage{soul}
\usepackage[colorlinks=true,urlcolor=brown,linkcolor=blue,citecolor=magenta]{hyperref}
\usepackage{float}
\usepackage{footmisc}
\usepackage{siunitx}

\usepackage[capitalise]{cleveref}

\definecolor{lightred}{rgb}{1,0.5,0.5}
\definecolor{lightgreen}{rgb}{0.5,1,0.5}
\definecolor{lightblue}{rgb}{0.5,0.5,1}
\definecolor{lightcyan}{rgb}{0.5,0.75,0.75}
\definecolor{lightmagenta}{rgb}{0.75,0.5,0.75}
\definecolor{customgreen}{rgb}{0.494,1,0.502}

\newcommand{\GeV}{~\mathrm{GeV}}

\newcommand{\Pdd}{\Phi_1^\dagger\Phi_1}
\newcommand{\Puu}{\Phi_2^\dagger\Phi_2}
\newcommand{\Pdu}{\Phi_1^\dagger\Phi_2}
\newcommand{\Pud}{\Phi_2^\dagger\Phi_1}

\begin{document}

\title{
Benchmarking a fading window: electroweak baryogenesis
in the C2HDM,\\[0.2em]
LHC constraints after Run~2 and
prospects for LISA
}

\author{Thomas Biekötter}
\email{thomas.biekoetter@desy.de}
\affiliation{
   Instituto de F\'isica Te\'orica UAM/CSIC,
   Calle Nicolás Cabrera 13-15,
   Cantoblanco, 28049, Madrid, Spain\\[0.4em]}

\author{María Olalla Olea-Romacho }
\email{maria_olalla.olea_romacho@kcl.ac.uk}
\affiliation{Theoretical Particle Physics and Cosmology, King’s College London, Strand, London WC2R 2LS, United Kingdom 
}

{\hfill\footnotesize IFT-UAM/CSIC-25-42}
\vspace*{1em}

\begin{abstract}
The origin of the baryon asymmetry of the universe
remains one of the most pressing open questions
in particle physics and cosmology.
Electroweak baryogenesis offers an experimentally testable
explanation, requiring new sources of CP violation
and a strong first-order electroweak phase transition.
The Two Higgs doublet model~(2HDM) is the simplest
scalar extension of the Standard Model that
can accommodate both ingredients.
We critically assess the viability of the complex 2HDM 
(C2HDM) (a 2HDM with a softly broken $\mathbb{Z}_2$
symmetry and a single source of explicit CP violation
in the Higgs sector) as a framework for electroweak baryogenesis,
incorporating for the first time a comprehensive
set of LHC Run~2 results at 13~TeV.
By defining
CP-violating benchmark
planes tailored for a strong first-order electroweak
phase transition, we identify regions of parameter space
motivated by electroweak baryogenesis
that will be testable at the LHC and at future
space-based gravitational wave experiments.
The benchmark planes are intended to
guide ongoing efforts in defining representative
scenarios for the exploration of CP-violation
in extended scalar sectors at the LHC Run~3
and beyond, while also assessing
the emerging synergy between the LHC and future gravitational wave
observatories such as LISA.
We also quantify the current tension
between the realisation of
electroweak baryogenesis
and the non-observation of the
electron electric dipole moment~(EDM),
finding that the predicted electron EDMs
typically exceed the experimental limits
by 
at least an order of magnitude.

\end{abstract}

\maketitle

\section{Introduction} 
\label{sec:intro}
The origin of the cosmic matter--antimatter asymmetry remains one of the most compelling open problems in fundamental physics. This asymmetry can be quantified in terms of the observed baryon-to-photon ratio~\cite{ParticleDataGroup:2024cfk}
\begin{equation}
\eta_\gamma \approx 6.14\times
10^{-10} \, .
\label{eq:bau}
\end{equation}
A dynamical explanation of this baryon asymmetry
of the universe~(BAU) requires the three Sakharov
conditions to be fulfilled: violation of baryon number,
violation of C and CP symmetry, and a significant
departure from thermal equilibrium~\cite{Sakharov:1967dj}.
While the Standard Model~(SM) contains ingredients that
satisfy the Sakharov conditions, such as baryon number
violating processes via electroweak~(EW)
sphalerons~\cite{Klinkhamer:1984di,Kuzmin:1985mm} and CP
violation through the complex
phase in the CKM matrix, it fails to generate a
sufficient BAU.
For a Higgs boson mass of 125~GeV, the EW phase
transition~(EWPT) is a smooth crossover rather
than a strong first-order EWPT~\cite{Kajantie:1996mn},
thereby failing the required
departure-from-equilibrium condition.
Furthermore, even under the presence of a sufficiently
strong EWPT, the CP violation contained in the
SM is far too small to produce the observed
BAU~\cite{Gavela:1993ts}.

The shortcoming of the SM to explain the BAU
has motivated many extensions of the Higgs sector aimed at realising EW baryogenesis~(EWBG)~\cite{Kuzmin:1985mm}.
The simplest and most studied scenario
that can realise EWBG is the
two Higgs doublet model~(2HDM)~\cite{Fromme:2006cm}.
By introducing a second Higgs doublet, the 2HDM
can substantially alter the EW symmetry-breaking
dynamics and provide additional CP-violating phases
beyond the CKM matrix in the Higgs sector.
It has long been known that 2HDMs can accommodate a
strong first-order EWPT~\cite{Cline:1996mga,
Ginzburg:2009dp,
Dorsch:2016nrg,Basler:2016obg,Dorsch:2017nza,
Bernon:2017jgv,Basler:2017uxn,
Su:2020pjw,
Basler:2021kgq,Goncalves:2021egx,
Enomoto:2022rrl, Goncalves:2023svb,Bittar:2025lcr, Olea-Romacho:2023rhh},
for example, through relatively large mass splittings
between the additional Higgs bosons predicted by
the model~\cite{Dorsch:2016nrg, Su:2020pjw, Biekotter:2022kgf,
Biekotter:2023eil}.
However, the parameter space that enables successful
EWBG in the 2HDM is increasingly under 
pressure from experimental 
observations.
The non-decoupling effects required
to generate a sufficiently strong EWPT prefer
a light spectrum of the additional Higgs bosons
not too far beyond the EW scale, making the EWPT
a collider target~\cite{Ramsey-Musolf:2019lsf}.
Accordingly, the non-observation of
beyond the SM~(BSM)
Higgs bosons at the Large Hadron collider~(LHC)
are able to exclude large parts of the
parameter space~\cite{Azevedo:2023zkg,Biekotter:2024ykp},
in particular the regions of parameter
space suitable for EWBG~\cite{Basler:2021kgq, 
Goncalves:2023svb}.

This is especially true for the searches performed
during the LHC Run~2 at an unprecedented center-of-mass
energy of 13~TeV, where novel signatures including
the production of more than one BSM
Higgs boson have been considered.
In this paper, we will
for the first time confront the
parameter regions favoured by EWBG in
the complex 2HDM~(C2HDM)~\cite{Weinberg:1990me,Ginzburg:2002wt},
the CP-violating 2HDM with softly-broken
$\mathbb{Z}_2$ symmetry,
with a comprehensive set of cross section
limits from LHC searches
performed at 13~TeV using the public
code \texttt{HiggsTools}~\cite{Bechtle:2008jh,
Bechtle:2011sb,Bechtle:2013wla,Bechtle:2020pkv,
Bahl:2022igd}.
This will allow us to define benchmark planes for
the C2HDM that are compatible with current LHC
constraints. These benchmark planes
are intended to guide the ongoing
efforts in defining representative scenarios for
the exploration of CP-violation in extended scalar
sectors at the LHC Run~3 and beyond.
We will focus on benchmark planes tailored for
a strong EWPT, identifying regions of parameter space
suitable for EWBG that will be
further probed 
in future Runs of the LHC.

In addition to assessing the experimental viability
of these regions in light of LHC Run~2 results
at $13\,\text{TeV}$, we examine whether they could
lead to complementary signals at future space-based
gravitational wave~(GW) detectors such as
LISA~\cite{Dorsch:2016nrg,Goncalves:2021egx,Biekotter:2022ckj}.
GW from a strong first-order EWPT are often highlighted
as a key signature of EWBG. We will show that there
are regions of parameter space that are compatible
both with EWBG and with GW signals that are 
detectable with LISA. However, in most parts
of the parameter space regions favoured by EWBG,
the predicted signal-to-noise ratios~(SNRs)
lie below LISA's
sensitivity threshold, posing a challenge for
detection~\cite{Goncalves:2023svb}.
Here it is important to note that the
theoretical uncertainties in the prediction of
the parameters characterizing the EWPT suffer from
sizable theoretical
uncertainties~\cite{Croon:2020cgk,Athron:2022jyi,Lewicki:2024xan}.
These uncertainties translate into substantial
uncertainties in the predictions for the BAU
and the GW signals,
such that the actual prospects for detection
could be somewhat more optimistic than our
estimates suggest.

Some of the benchmark planes that we are proposing
are motivated by studies of the EWPT performed in the
real 2HDM~(R2HDM), where the Higgs sector
is assumed to conserve the CP symmetry.
Studies in the R2HDM have shown that 
a large separation between a heavier new
CP-odd Higgs boson $A$ and a lighter new CP-even
Higgs boson $H$ facilitates
the strongest EWPTs~\cite{Basler:2016obg, Biekotter:2022kgf,
Biekotter:2023eil}. An important consequence is
that this mass hierarchy allows kinematically for
the decay $A \to ZH$.
This signature has consequently been
dubbed a ``smoking gun'' signature of a strong
first-order EWPT in the 2HDM~\cite{Dorsch:2014qja, 
Dorsch:2016nrg, Biekotter:2023eil}
since the same physics (a large coupling and mass gap)
that facilitates a strong EWPT, also increases
the cross sections for the $A \to ZH$
signal~\cite{Biekotter:2023eil, Arco:2025ydq}.
We will demonstrate here with a selection of
benchmark scenarios that
the C2HDM inherits this
phenomenology, with sizable mass gaps
between the BSM neutral scalars facilitating
the strongest transitions, and we study the
impact of the CP violation in the Higss sector
on the strength of the EWPT in such scenarios.
Moreover, we will analyse which additional collider
signatures play a role in the C2HDM that can probe
parameter regions with a strong EWPT, which are not
present in the R2HDM.

We also present a benchmark plane in which
a strong EWPT is realised without large mass
splittings between the BSM scalars, such that
the decay $A \to ZH$ is absent.
This possibility requires small values of $\tan\beta$
(defined as the ratio of the vacuum expectation
values~(vev) of the two Higgs doublets) and an
overall lighter spectrum~\cite{Basler:2017uxn,Dorsch:2017nza}.
While we give an explicit example of such a scenario
that remains compatible with existing limits
from LHC searches,
we will argue that this possibility is on
the verge of being excluded.
Accordingly, there are good
prospects for finally excluding (or confirming,
e.g.~via the detection of the charged scalars)
this scenario during Run~3.
We also find that the
strength of the EWPT in almost
mass degenerate scenarios is substantially smaller,
barely satisfying the baryon number preservation
condition~\cite{Dimopoulos:1978kv,Kuzmin:1985mm}
that prohibits the washout of the
BAU after the transition.
In addition, the smaller strengths of the EWPT
in this scenario gives rise to GWs with
LISA SNRs of the order of~$10^{-8}$ or below,
such that it is unlikely that it can be probed with LISA.

Another key challenge for EWBG that has emerged in
recent years is that scenarios with new sources
of CP violation are stringently constrained by
experimental limits on electric dipole
moments~(EDMs), in particular the electron
EDM~\cite{ACME:2018yjb,Roussy:2022cmp}.
In the C2HDM, the same CP-violating phase responsible
for generating the BAU also induces an electron
EDM at the two-loop level via Barr–Zee-type diagrams
that typically is substantially
larger than the experimental
upper bounds~\cite{Biekotter:2024ykp}.
As such, the non-observation of an EDM places
strong pressure on the parameter space relevant
for EWBG. In this work, we compute the predicted
electron EDMs across our benchmark scenarios,
but we do not impose the experimental bounds as
a hard constraint. Instead, we indicate the
electron EDM values on our benchmark planes
in order to quantify the level of tension.
This approach reflects the possibility that the
electron EDM may be suppressed in more general
extensions of the C2HDM, e.g.~via additional
CP-violating phases (if hard breaking of the $\mathbb{Z}_2$
symmetry is considered)~\cite{Enomoto:2022rrl}
or novel mechanisms such as
secluded~\cite{Cline:2017qpe}
or transient~\cite{Huber:2022ndk}
CP violation, without affecting the
dynamics of the EWPT.

Taking into account both the new LHC Run~2 results
at 13~TeV and the non-observation of the electron EDM,
our findings suggest that the C2HDM may face
challenges as a benchmark model for EWBG.
In addition, only a small part of the parameter region
suitable for EWBG produces GW signals that are strong
enough to be detectable with LISA.
Nevertheless, our benchmark planes reveal that
interesting parameter regions still persist,
some of which can serve as well-motivated benchmark
scenarios for future LHC searches,
and a subset even includes regions where successful
EWBG is accompanied by GW signals within LISA's
projected sensitivity.

\section{The 2HDM}

In this section, we briefly introduce the 2HDM and
establish our notation, beginning with the
CP-conserving R2HDM.
Although our primary focus lies on the CP-violating
C2HDM, starting from the CP-conserving case
allows us to point out more transparently the
differences in the
dynamics of the EWPT and the role of CP violation
in our numerical analysis.

\subsection{The real two-Higgs-doublet model}
\label{sec:real2HDM}

In this section, we present a short overview of the CP-conserving R2HDM. 
The Higgs sector
consists of two SU(2) Higgs doublets, $\Phi_1$ and $\Phi_2$, both carrying
hypercharge $1/2$.\footnote{For a
comprehensive review of the 2HDM, see
e.g.~Ref.~\cite{Branco:2011iw}.}
The doublets can be written in terms of their
neutral and charged components
\begin{equation}
\Phi_1 = 
\begin{pmatrix}
\phi_1^+ \\
(\phi_1 + i \, \sigma_1) / \sqrt{2}
\end{pmatrix},
\qquad
\Phi_2 = 
\begin{pmatrix}
\phi_2^+ \\
(\phi_2 + i \, \sigma_2) / \sqrt{2}
\end{pmatrix} \, .
\label{eq:fields_r2hdm}
\end{equation}
To suppress tree-level flavor-changing neutral currents, we impose a softly broken $\mathbb{Z}_2$ symmetry~\cite{Fayet:1974fj, Fayet:1974pd}
under which the Higgs doublets transform as
$\Phi_1 \to \Phi_1$ and $\Phi_2 \to -\Phi_2$,
and which extends to the Yukawa sector, see below.
At tree level, the scalar potential takes the form
\begin{align}
\label{eq:HiggsPotential_r2hdm}
& V_{\text{tree}}(\Phi_1,\Phi_2) = \\
& m_{11}^2\,\Pdd + m_{22}^2\,\Puu - \left(m_{12}^2\Pdu + \mathrm{h.c.}\right) \nonumber\\
&\hphantom{=} + \frac{1}{2}\lambda_1 (\Pdd)^2 + \frac{1}{2}\lambda_2 (\Puu)^2  + \lambda_3 (\Pdd)(\Puu) \nonumber        \\
&\hphantom{=} + \lambda_4 (\Pdu)(\Pud) + \frac{1}{2} \left(\lambda_5(\Pdu)^2 + \mathrm{h.c.}\right) \, . \nonumber 
\end{align}
where all parameters are real, ensuring CP-conservation. The vacuum expectation values (vevs) of the Higgs doublets are given by $\langle \phi_1 \rangle = v_1$ and $\langle \phi_2 \rangle = v_2$, satisfying the relation $v_1^2 + v_2^2 \equiv v^2 \approx 246\GeV$, with the ratio of vevs parameterized as $\tan\beta \equiv t_{\beta} = v_2/v_1$. 

After electroweak symmetry breaking, the R2HDM predicts five physical Higgs states: two neutral CP-even scalars $h$ and $H$, one neutral CP-odd pseudoscalar $A$, and a pair of charged Higgs bosons $H^{\pm}$. The transition from the gauge basis to the mass basis is governed by two mixing angles: $\alpha$ for the CP-even sector and $\beta$ for the CP-odd/charged sector.
The lighter CP-even state $h$ is conventionally
identified as the observed Higgs boson~\cite{Aad:2012tfa,
Chatrchyan:2012ufa} at $m_h = 125\GeV$.
The \textit{alignment limit} is defined by~\cite{Gunion:2002zf}
\begin{equation}
\cos(\beta-\alpha) = 0 \, ,
\label{eq:alignlim_r2hdm}
\end{equation}
where the couplings of $h$ match those of the SM
Higgs boson at tree level.

Since $\Phi_1$ and $\Phi_2$ transform differently under the $\mathbb{Z}_2$ symmetry, each fermion type can only couple to one of the Higgs doublets. This results in four different 2HDM types, depending on how the $\mathbb{Z}_2$ symmetry is assigned in the Yukawa sector.
For our benchmark planes, we focus on the type~I
where all fermions couple to the Higgs doublet
$\Phi_2$, as it is the least constrained by
current LHC data among the four types.
In particular, it allows for the largest deviations
from the alignment limit without conflicting with
measurements of the 125~GeV Higgs 
boson~\cite{Biekotter:2022ckj,Bahl:2022igd}.
Notably, in the exact alignment limit,
the direct production of the additional neutral and
charged scalars is suppressed with increasing $t_\beta$,
thereby evading many of the existing constraints
from searches for these particles.

Although our analysis is performed within the C2HDM,
it is useful to define the set of input parameters
of the R2HDM, as this facilitates a transparent
comparison in the CP-conserving limit where
(in our parameter basis of the C2HDM, see below)
the C2HDM smoothly reduces to the real 2HDM.
The 2HDM parameter space
can be most conveniently investigated with the following
set of input parameters,
\begin{equation}
t_{\beta}, \; M, \; v, \; \cos (\beta - \alpha), \; m_h, \; m_H, \; m_A, \; m_{H^{\pm}} \, ,
\label{eq:input_parameters_r2hdm}
\end{equation}
where we defined the $\mathbb{Z}_2$-breaking scale
\begin{equation}
  M^2 = \frac{m_{12}^2}{s_\beta c_\beta} \, ,
\end{equation}
and where $v \approx 246\GeV$ and
$m_h \approx 125\GeV$ are fixed from experimental input.

\subsection{The complex two-Higgs-doublet model}
\label{sec:complex2HDM}

Analogous to the R2HDM, the tree-level scalar
potential of the C2HDM is constructed from two
$\rm{SU(2)}_L$ scalar doublets, $\Phi_1$ and $\Phi_2$,
both carrying a hypercharge of $1/2$, as shown
in \cref{eq:fields_r2hdm}.
The scalar potential has the same form as in
\cref{eq:HiggsPotential_r2hdm}, respecting a $\mathbb{Z}_2$
symmetry only softly broken by the term proportional
to $m_{12}^2$,
but now allowing $m_{12}^2$ and $\lambda_5$ to take
complex values~\cite{Weinberg:1990me,
Ginzburg:2002wt}.
Without loss of generality, we
work in a basis in which at zero
temperature the vevs
of the neutral components of the Higgs doublets
are real and positive numbers, i.e.~$\langle \phi_1
\rangle = v_1$ and $\langle \phi_2 \rangle = v_2$
\cite{Ginzburg:2009dp}.
This promotes
$t_\beta = v_1 / v_2$ to a physical parameter
parameter once a specific Yukawa type is
chosen~\cite{Boto:2020wyf}.
For the discussion of the EWPT
in \cref{sec:ThermalHistory}, we note that
at finite temperature $T$, the Higgs doublets can
develop CP-violating and charge-breaking vevs
$v_{\rm CP}$ and $v_{\rm CB}$, respectively.
Then, the most general vacuum can be written as
\begin{equation}
\langle \Phi_1(T) \rangle = 
\begin{pmatrix}
0 \\
v_1(T)
\end{pmatrix} , \,
\langle \Phi_2(T) \rangle = 
\begin{pmatrix}
v_{\rm CB}(T) \\
v_2(T) + i \, v_{\text{CP}}(T)
\end{pmatrix} \, ,
\label{eq:vevs}
\end{equation}
with $v_{\rm CP}(T = 0) = v_{\rm CB}(T = 0) = 0$.
We included the possibility of charge-breaking
phases in our analysis of the thermal
history at finite temperature,
since a charge-breaking epoch
during or after the EWPT would invalidate
our computation of the BAU.
However, we only
encountered phase with $v_{\rm CB} \neq 0$,
and no transient charge-breaking occurred.

In contrast to the R2HDM,
the C2HDM allows for explicit
CP violation through complex parameters
in the scalar potential.
Making use of a global rephasing invariance and
the freedom to perform phase rotations of the Higgs
doublets~\cite{Ginzburg:2002wt,Ogreid:2018bjq},
only one independent complex phase remains.
All parameters can be taken real except for
the soft $\mathbb{Z}_2$-breaking term $m_{12}^2$
and the quartic coupling $\lambda_5$,
whose complex phases are related by
\begin{equation}
\mathrm{Im}(m_{12}^2) = \frac{1}{2}s_\beta c_\beta v^2
 \mathrm{Im(\lambda_5)} \, ,
 \label{eq:imm12sq}
\end{equation}
expressing the fact that there is only one
independent CP-violating phase in the model.

The physical spectrum after EW symmetry breaking
consists of three neutral scalars $h_1$, $h_2$, and $h_3$, which are, in general, admixtures of CP eigenstates. In what follows, we assume a mass ordering given by $m_{h_1} = 125~\mathrm{GeV} < m_{h_2} < m_{h_3}$, along with a pair of charged Higgs bosons $H^\pm$ with mass $m_{H^\pm}$.
The rotation from the 
interaction basis
$\phi_1$, $\phi_2$ 
and $A = -s_\beta \sigma_1
+ c_\beta \sigma_2$ 
to the mass eigenstate
basis $h_{1,2,3}$ 
can be expressed via a $3\times3$ orthogonal
rotation matrix $R_{ij}$,
which we
parametrise using the
three mixing angles $\alpha_1$, $\alpha_2$, and $\alpha_3$,
\begin{equation}
R =
\begin{pmatrix}
c_1 c_2 & s_1 c_2 & s_2 \\
- (c_1 s_2 s_3 + s_1 c_3) & c_1 c_3 - s_1 s_2 s_3 & c_2 s_3 \\
- c_1 s_2 c_3 + s_1 s_3 & - (c_1 s_3 + s_1 s_2 c_3) & c_2 c_3
\end{pmatrix} \, ,
\label{eq:rmatrix}
\end{equation}
with the shorthand notation
$s_i = \sin\alpha_i$, $c_i = \cos\alpha_i$,
$(i = 1,2,3)$.
The admixture of the states $\phi_{1,2}$ with
the state $A$ signals CP violation, whereas in the
CP-conserving limit, e.g.~$\alpha_2 = \alpha_3 = 0$,
the mixing matrix $R$ decomposes into a block diagonal
form with a $2 \times 2$ matrix rotating the fields
$\phi_{1,2}$ into two CP-even mass eigenstates,
and the field $A$ becomes a CP-odd mass eigenstate.
As discussed above,
there is only one independent CP-violating 
phase in the C2HDM.
However, both angles $\alpha_2$
and $\alpha_3$ parametrise the CP-violating mixing in
the neutral scalar sector.
As a consequence, in the C2HDM it is not possible
to use all physical scalar masses and all mixing
angles as independent input parameters, in contrast
to the real 2HDM as shown
in \cref{eq:input_parameters_r2hdm}. 
Different strategies have been
adopted in the literature to define a set of input
parameters. One common approach is to use all three
mixing angles $\alpha_{1,2,3}$ as free parameters, in which
case only two of the three neutral scalar masses
can be kept as free parameters, and one is a dependent
parameter (see,
e.g.~Refs.~\cite{Basler:2017uxn,Fontes:2017zfn,
Basler:2021kgq,
Azevedo:2023zkg,
Biekotter:2024ykp}). However,
this parametrisation makes it impossible to study
benchmark planes with all physical masses kept fixed,
which is useful to isolate the impact of CP violation
on the dynamics of the EWPT, which we want to
investigate here.
Moreover, in general it does not allow to continuously
approach the CP-conserving limit because the predicted
mass can become tachyonic.

We therefore use a different parametrisation in
which only the two mixing angles $\alpha_1$ and
$\alpha_3$ are independent input parameters,
whereas $\alpha_2$ is a derived parameter. Similar
parameterizations of the C2HDM with slightly
different notations
have been used in the past
in the context of EWBG in Refs.~\cite{Shu:2013uua,
Inoue:2014nva,Chen:2015gaa,
Goncalves:2023svb}.
In total, we use the following set of input
parameters to define our benchmark planes,
\begin{equation}
t_{\beta}, \; M, \; v, \; \alpha_{1}, \; \alpha_3, \;
m_{h_{1}}, \; m_{h_2}, \; m_{h_3}, \; m_{H^{\pm}} \, ,
\label{eq:input_parameters_c2hdm}
\end{equation}
where the $\mathbb{Z}_2$-breaking scale in the
C2HDM is defined by
\begin{equation}
  M^2 = \frac{\mathrm{Re}(m_{12}^{2})}{s_\beta c_\beta} \, .
  \label{eq:M}
\end{equation}
The set of equations that relates this set of
parameters to the parameters of the potential
shown in \cref{eq:HiggsPotential_r2hdm} are given
in \cref{app:pararelations}.
In this parametrisation, 
the angle $\alpha_1$ acts as a parameter
quantifying departures from the
C2HDM alignment limit.
Under the presence of small CP-odd components
in the SM-like Higgs boson $h_1$
(e.g.~$\alpha_3 \approx 0$, see discussion below)
as required in
the Yukawa type~I to be compatible with
LHC data~\cite{Azevedo:2023zkg}, the alignment limit
in the C2HDM can be approximately defined by
\begin{equation}
  s_{\beta - \alpha_1} \equiv
  \sin (\beta - \alpha_1) \approx 0 \, ,
  \label{eq:alignlim_c2hdm}
\end{equation}
such that $\alpha_1 \neq \beta$ signal departures
from the alignment limit.\footnote{Due
to different
conventions for defining the rotation matrices,
the alignment limit in the C2HDM is defined with
the sine, whereas the alignment limit in the
real 2HDM is typical defined with the cosine,
see \cref{eq:alignlim_r2hdm}.}
The angle $\alpha_3$ acts as a parametrisation of the amount of CP violation.
The CP-conserving limit is achieved for either
$\alpha_3 \to 0$,
for which $h_2$ becomes a pure CP-even state
corresponding to the state~$H$ of the real
2HDM and $h_3$ becomes a CP-odd state corresponding
to the state~$A$ of the real 2HDM,
or $\alpha_3 \to \pm \pi / 2$,
for which conversely $h_2$ and $h_3$ correspond
to the states $A$ and $H$, respectively.
In both cases,
the derived parameter $\alpha_2$ vanishes.

\subsection{Theoretical and experimental constraints}
\label{sec:constraints}

All benchmark planes defined in this work satisfy
the following set of theoretical and
experimental constraints.

\subsubsection{Theoretical constraints}
\label{sec:theoconstraints}

On the theory side, we ensure absolute stability
of the EW vacuum for all points featuring
a strong EWPT using the public
code \texttt{BSMPT v.3}~\cite{Basler:2018cwe,
Basler:2020nrq,Basler:2024aaf}.
To this end, we verify that the effective
potential at one-loop level and zero temperature
(see the discussion in \cref{sec:ThermalHistory})
is bounded from below and that the minimum
corresponding to the EW vacuum is the global
minimum of the potential.
Near the alignment limit, the
bounded-from-below conditions are most naturally satisfied
by requiring $M \leq m_{h_2}, m_{h_3}, m_{H^\pm}$, a condition we impose throughout to define the benchmark planes.

We note that, for a subset of parameter points
featuring a strong EWPT and a
stable EW
vacuum at one-loop level, the tree-level scalar potential
is unbounded from below. While this does not constitute
an inconsistency (since quantum corrections can indeed
stabilise or destabilise the potential)
it implies that the vacuum stability of these points
relies on the one-loop contributions.
This introduces a degree of theoretical uncertainty,
as corrections from higher orders could, in principle,
alter the stability of the vacuum again.
We find that the boundary between parameter space
regions with a bounded and unbounded potential
shifts only slightly between the tree-level and
one-loop analyses.
To maintain clarity in our presentation, we do not indicate
tree-level bounded-from-below exclusions in the plots,
as this would in general be too restrictive~\cite{Staub:2017ktc}.
Importantly, we have verified that the vast majority
of these points lie in regions already excluded by
current LHC searches, as discussed in \cref{sec:lhc},
such that this issue does not have an impact on
our conclusions regarding the possibility to realise
EWBG or predict GW signals potentially detectable
with LISA.

To ensure perturbativity, we use the public code
\texttt{ScannerS v.2}~\cite{Muhlleitner:2020wwk} to test 
each parameter point against leading-order perturbative unitarity 
constraints in the high-energy 
limit~\cite{Kanemura:1993hm},
ensuring that, if not indicated otherwise,
the whole benchmark planes lie within the 
perturbative regime.
These conditions place upper bounds on the absolute values of the quartic scalar couplings $\lambda_i$, which govern both the mass splitting between the $\mathbb{Z}_2$-breaking scale $M$ and the BSM scalar masses, as well as the mass differences among the BSM states themselves. As a result, perturbative unitarity
conditions exclude parameter regions where
the difference among the parameters
$M$, $m_{h_2}$, $m_{h_3}$ and $m_{H^\pm}$ are
substantially larger than $v \approx 246\GeV$.

\subsubsection{EW precision observables}

On the experimental side, we make sure that our
benchmark planes are compatible with constraints
from EW precision observables in the
form of the oblique parameters $S$, $U$ and
$T$~\cite{Peskin:1991sw}.
Close to the alignment limit and with small
CP-violating mixing in the scalar sector,
these constraints can be satisfied if at least
one of the BSM neutral state $h_2$ or $h_3$
is approximately mass degenerate with the
charged scalars $m_{H^\pm}$~\cite{Gerard:2007kn}.
Under the presence of large CP-violating
mixing between $h_2$ and $h_3$, the constraints
on the EW precision observables can be most easily
satisfied if
all BSM states are approximately mass degenerate. 

\subsubsection{Flavour-physics observables}

Further indirect constraints on the C2HDM
parameter space arise from flavour-physics
observables. In particular, measurements of
radiative $B$-meson decays in the form
of $\bar B \to X_s \gamma$ transitions
and of leptonic $B$-meson decays from
$B^0 \to \mu^+ \mu^-$ decays provide fairly
robust lower limits on $t_\beta$.
We have verified using the public code \texttt{thdmTools v.1.3}~\cite{Biekotter:2023eil}, interfaced with \texttt{SuperIso v.4.1}~\cite{Mahmoudi:2008tp, Neshatpour:2021nbn}, that current experimental results~\cite{Altmannshofer:2021qrr,CMS:2022mgd,LHCb:2021awg} exclude values of $t_\beta \lesssim 1.5$ at approximately the $2\sigma$ level for charged Higgs boson
masses in the range $400\GeV \lesssim m_{H^\pm} \lesssim 800\GeV$, across all Yukawa types. Thus, for our benchmark
planes in type~I we use values of $t_\beta \geq 1.5$.
For the type~II and the flipped type,
the constraints from $\bar B \to X_s \gamma$ transitions
give rise to an additional lower limit
of $m_{H^\pm} \gtrsim 500\GeV$~\cite{Azevedo:2023zkg,
Misiak:2017bgg}.\footnote{Three
of the four benchmark planes presented below
feature values of $m_{H^\pm} \geq 650\GeV$, such that
they satisfy the constraints from $\bar B \to X_s \gamma$ 
transitions in all Yukawa types and not only in the
type~I considered here.
The final benchmark plane with an almost mass
degenerate spectrum and $m_{H^\pm} = 400\GeV$ would
be in tension with this constraint in type~II
and the flipped Yukawa type.}

\subsubsection{LHC searches}
\label{sec:lhc}

Regarding collider constraints, we incorporate the
latest LHC results from searches for additional
Higgs bosons and from cross-section measurements
of the 125~GeV Higgs boson using the public tool
\texttt{HiggsTools v.~1.2}~\cite{Bahl:2022igd}.
In this way we can include for the first time a
comprehensive set of LHC Run~2 results
at~13~TeV in an analysis of the EWPT
in the C2HDM, while in
previous studies of the EWPT in
the C2HDM (and extensions thereof)
the wealth of new final states explored during
Run~2 was not yet included~\cite{Goncalves:2023svb,
Basler:2021kgq,Aiko:2025tbk,Athron:2025iew}.\footnote{The C2HDM analyses
from Refs.~\cite{Goncalves:2023svb,
Basler:2021kgq} relied on the codes \texttt{HiggsBounds v.5}
and \texttt{HiggsSignals v.2}
which were deprecated (in favour of \texttt{HiggsTools})
and last updated with new experimental data in 2021,
at which point the major part of LHC results based
on the full Run~2 dataset was not yet available.}
\texttt{HiggsTools} enables us to
include a wide range of dedicated searches for
additional Higgs bosons, many of which are
particularly sensitive to CP-violating scenarios.
Since our goal is to provide benchmark planes
as a road map for future LHC searches, it is crucial
to verify that the corresponding parameter
regions are compatible with the already
existing limits.

LHC searches that become especially relevant
under the presence of CP violation in the C2HDM
include searches for a heavy spin-0
resonance $X$ decaying into $Z h$, where $h$
is SM-like Higgs boson at
125~GeV~\cite{CMS:2019qcx,ATLAS:2022enb}.
While in the C2HDM the state
$X$ can be either $h_2$ or $h_3$
(if kinematically allowed),
in the CP-conserving 2HDM only the decay
$A \to Z h$ is allowed at tree level, whereas
$H \to Z h$ is forbidden.
The presence of two scalars, $h_2$ and $h_3$, potentially
decaying into the 125~GeV Higgs boson and
a $Z$-boson makes the $Zh$ final state exceptionally
important~\cite{Chen:2015gaa,Fontes:2015xva,Azevedo:2023zkg}.
We will also discuss the impact of these
searches in comparison to the smoking-gun signature
$A \to Z H$ from the R2HDM, with the heavier CP-even BSM
state $H$ in the final state, see the discussion
in \cref{sec:intro}. 
This process and its role
in probing the EWPT in the R2HDM has been
extensively discussed in previous
works~\cite{Goncalves:2022wbp,Biekotter:2022kgf,Biekotter:2023eil,
Arco:2025ydq}, and we will
here focus on its impact in the C2HDM.

In addition, the increased mass reach of Run~2
at 13~TeV allows for probing a richer Higgs-boson
pair-production phenomenology.
Searches for the pair-production of the 125~GeV
Higgs boson have become a central part of
the future LHC Higgs-physics
programme~\cite{CMS:2018ipl,ATLAS:2023vdy}.
In the C2HDM, $h_1$-pairs can be resonantly
produced by both $h_2 \to h_1 h_1$ and
$h_3 \to h_1 h_1$ decays if kinematically allowed.
On the other hand, the CP-conserving 2HDM only
allows for the decay $H \to hh$, whereas
$A \to hh$ is forbidden.
With the 13~TeV data collected during Run~2,
ATLAS and CMS also reported for the first time
results from searches for a heavy spin-0 resonance
$X$ decaying into a SM-like Higgs boson and
another lighter new spin-0
resonance~$Y$~\cite{CMS:2021yci,
ATLAS:2021ldb}.\footnote{There
are further very recent results by
CMS~\cite{CMS:2023boe,CMS-PAS-B2G-24-001}
and ATLAS~\cite{ATLAS:2024xkk}
that are not yet included in \texttt{HiggsTools}
and therefore not taken into account here.}
If kinematically allowed, in the C2HDM this
$X \to h Y$ channel is in general realised
via $h_3 \to h_1 h_2$ decays irrespective of
the precise CP-admixtures of the states~$h_i$.
On the contrary, there is no $hHA$-coupling
in the R2HDM,
such that searches for $X \to h Y$ with
$m_Y \neq 125\GeV$ cannot probe the R2HDM.
In addition to becoming more relevant in the
presence of CP violation~\cite{Fontes:2017zfn},
such multi-scalar signatures are also
increasingly important in the context of EWPTs,
as they are governed by triple-scalar
couplings that directly probe the shape of the
scalar potential, potentially providing
crucial information about the dynamics of the transition.

In addition to the novel signatures discussed above,
also the traditional search channels for a Higgs boson
decaying (depending on the mass range) into pairs of
massive gauge bosons or third-generation fermions
are affected by the presence of CP violation.
For example, in the R2HDM the CP-odd state~$A$
does not couple at tree level to gauge bosons,
such that its decays into $WW$ and $ZZ$ are highly
suppressed, and only the decays $H \to WW$ and
$H \to ZZ$ are promising. In the C2HDM, outside of
the alignment limit both $h_2$ and $h_3$ can decay
into massive gauge bosons with sizable rates,
giving rise to stronger constraints
especially for masses
below the di-top threshold~\cite{Keus:2015hva}. Finally,
above the di-top
threshold, searches for heavy Higgs bosons decaying
into $t \bar t$ pairs exclude small $t_\beta$-values
of $t_\beta \approx 1$ in the CP-conserving
2HDM~\cite{CMS:2019pzc}.
A CP-violating mixing between heavy states $h_2$
and $h_3$ can yield interesting signal-signal
interference patterns
in the $m_{t \bar t}$ invariant mass
distributions~\cite{Bahl:2025you}
which have not been explored experimentally yet.
Instead, in the C2HDM, the low-$t_\beta$ region can
currently constrain more robustly in final states
with four top quarks~\cite{CMS:2019rvj,ATLAS:2024jja},
which is much less sensitive to signal-background
and signal-signal interference effects.

Finally, it is known that the regions with a strong
EWPT in the 2HDM are correlated with sizable
radiative enhancements of the 125~GeV Higgs boson
self-coupling by about
50~to~150\%~\cite{Kanemura:2004ch,Basler:2017uxn}.
In particular, parameter points
featuring an EWPT strong enough to source
a GW signal potentially detectable at LISA 
predict values of
$\kappa_\lambda \approx 2$~\cite{Biekotter:2022kgf},
where $\kappa_\lambda$ is the $hhh$-coupling normalized
to the SM leading-order prediction.
At the LHC, the Higgs boson self-coupling is directly
accessible via non-resonant Higgs boson pair production.
Due to destructive interference effects, the
enhancements of the self-coupling yield a suppression
of the total cross section for $hh$-production at
the LHC. Current upper limits on $\kappa_\lambda$
are at the level of $\kappa_\lambda \lesssim 7$ at
95\% confidence level~\cite{ATLAS:2024ish,CMS:2024awa}.
Thus, the measurements of the
Higgs boson self-coupling are currently still far from probing
models featuring a strong EWPT.
We therefore do not
focus on constraints on the Higgs boson self-coupling
in our discussion. However, the projected sensitivity
from the high-luminosity LHC indicates that it
will be possible to set a limit of $\kappa_\lambda < 1.6$
at 95\% confidence level~\cite{CMS:2025hfp}.
Accordingly, during the high-luminosity stage,
the LHC is expected to be able to
probe the parameter space regions with the strongest
EWPT in the 2HDM via non-resonant
Higgs boson pair production.

\subsubsection{Electron EDM}
\label{sec:edm}

New sources of CP violation can be constrained
with measurements of EDMs.
For simplicity, we focus exclusively on the
electron EDM, as it provides the most robust
and model-independent constraint on new
CP-violating phases in the Higgs sector.
In our analysis, we compute the EDM of the
electron at the two-loop
level using Barr–Zee-type diagrams~\cite{Abe:2013qla},
as implemented in the public code
\texttt{ScannerS v.2}~\cite{Muhlleitner:2020wwk}.
In the absence of suppression mechanisms,
heavy new physics with CP-violating interactions
generates an electron EDM that
roughly scales via~\cite{Morrissey:2012db}
\begin{equation}
  |d_e| \sim \frac{m_e}{1~\mathrm{MeV}}
    \left( \frac{1~\mathrm{TeV}}{M_{\rm UV}} \right)^2
    \cdot 10^{-26}~e~\mathrm{cm} \, ,
  \label{eq:scaling_edm}
\end{equation}
where $M_{\rm UV}$ represents the mass scale of the particles
whose interactions are CP-violating, and
$m_e$ is the electron mass.
In the C2HDM, with new Higgs bosons at the
EW scale, one therefore expects to find values
of the order of $|d_e| \sim 10^{-28} e  \, \mathrm{cm}$.
The theoretical predictions can
be compared to the current
experimental upper bounds set by the ACME
collaboration in 2018~\cite{ACME:2018yjb},
\begin{equation}
  |d_e| < 1.1 \cdot 10^{-29} e \, \mathrm{cm} \quad
  \textrm{at 90\% CL} \, ,
\end{equation}
and the JILA collaborations
in 2022s~\cite{Roussy:2022cmp},
\begin{equation}
  |d_e| < 4.1 \cdot 10^{-30} e \, \mathrm{cm} \quad
  \textrm{at 90\% CL} \, ,
\end{equation}
These are the most stringent limits
on new sources of CP violation to date.
The experimental upper limits are about two orders
of magnitude below the values suggested by
the scaling argument discussed above,
demonstrating the current tension between
C2HDM predictions and
the non-observation of an electron EDM.
In principle, it is possible to fine-tune the parameters of the C2HDM so that different two-loop contributions to the electron EDM interfere destructively, thereby reducing the predicted value below the current experimental bounds~\cite{Fontes:2017zfn,Azevedo:2023zkg}.
One could then investigate whether these finely
tuned parameter points allow for a strong EWPT.
However, this approach relies on substantial cancellations that are highly sensitive to the choice of renormalization scheme for the fermion masses (e.g., $\overline{\textrm{MS}}$ vs.\ on-shell)~\cite{Azevedo:2023zkg}, and are generally expected to be unstable under higher-order corrections.
Consequently, this strategy introduces significant
theoretical uncertainty, rendering such parameter
points unreliable for drawing firm conclusions
about the viability of EWBG in the C2HDM.

We here follow a different approach.
Rather than imposing the experimental limits as
a strict constraint on the parameter space,
we indicate the predicted values across our
benchmark planes. This facilitates a direct and
quantitative comparison between theoretical predictions
and experimental bounds, making it possible to
assess the level of tension in regions relevant
for successful EWBG.
There are compelling reasons
to quantify this tension, rather than eliminate
all C2HDM parameter points which do not satisfy
the limit on the electron EDM.
In the C2HDM, the electron EDM and the BAU both
arise from the same unique source of CP violation.
This tight connection enhances predictability but
also means that EDM constraints strongly impact the
viability of the C2HDM as a framework for EWBG.
However, this tight correlation can be relaxed
in more general frameworks.

For example, introducing a hard breaking of the
$\mathbb{Z}_2$ symmetry imposed in the
C2HDM allows for additional
CP-violating phases in the Higgs potential
and the Yukawa sector.
Regarding the latter, a possibility to have
CP-violating phases while suppressing flavour-changing
neutral currents consists of the so-called
flavour alignment~\cite{Pich:2009sp,Jung:2013hka}.
Such models can suppress the predicted electron
EDM~\cite{Azevedo:2023zkg,Banik:2024ugs,
Davila:2025goc}
while maintaining the CP violation required
for EWBG~\cite{Fuyuto:2019svr,Kanemura:2020ibp,
Enomoto:2021dkl,Enomoto:2022rrl}.
It is important to note, however,
that this approach comes with several caveats:
(1) a significant fine-tuning is required between
different and, a-priori, unrelated sources of CP
violation to suppress the EDMs,
(2) the suppression is generally only possible for
one specific fermion EDM, leaving other EDMs
unsuppressed,
(3) the hard $\mathbb{Z}_2$-breaking
in the Higgs sector induces
flavor-changing neutral currents that
are tightly constrained, and (4)
the flavor alignment is not stable under
the running group evolution,
introducing further naturalness issues.

Alternative but theoretically more elegant approaches
involve the introduction of additional fields.
For instance, the addition of gauge-singlet scalars
or further scalar doublets forming a dark/hidden
sector can lead to the so-called \textit{secluded}
CP violation, where CP-violating interactions are
confined to couplings involving dark-sector
particles~\cite{Cordero-Cid:2016krd,Cline:2017qpe,
Carena:2018cjh,Hall:2019ank}.
Then, EDMs are generated only at higher loop orders
beyond the two-loop level relevant in the C2HDM.
This setup can suppress EDMs to acceptable levels,
but it typically involves more parameters,
less predictivity, and potentially additional constraints
from dark matter phenomenology.
Another elegant idea is the so-called
\textit{transient} CP violation, where new
CP-violating sources are active only during the
high-temperature epoch of the early universe
and vanish at zero temperature~\cite{Cline:2012hg,
Inoue:2015pza,
Huber:2022ndk}.
This naturally suppresses EDMs today but requires
more complex models to dynamically realise a period
of transient CP violation in the early universe.

Lastly, some proposals~\cite{AharonyShapira:2021ize,
Bahl:2022yrs,Aiko:2025tbk}
aim to suppress the electron Yukawa coupling,
thereby reducing the leading contributions to
the electron EDM, which then arise only at
higher loop orders. While this can potentially
bring the EDM prediction below current
experimental limits, such suppression is
typically ad hoc and challenging to realise
within a UV-complete framework~\cite{Panico:2018hal}.
Moreover, this approach does not address EDM contributions
for other fermions or composite particles
(such as neutrons or molecules), which remain
sensitive to CP-violating effects and may yield
constraints.

\subsection{Thermal history and description
of the EWPT}
\label{sec:ThermalHistory}

To study the dynamics of the EWPT
in the C2HDM, we employ the public
code~\texttt{BSMPT v.3}~\cite{Basler:2018cwe,Basler:2020nrq,
Basler:2024aaf},
which computes the one-loop daisy-resummed finite-temperature effective potential in the Landau gauge
and using an on-shell-like
renormalization scheme.
In this approach, 
the total effective potential is given by
\begin{align}
V(\Phi_i,T) &= V_{\rm tree}(\Phi_i) +
  V_{\rm CW}(\Phi_i) + V_{\rm CT}(\Phi_i) \nonumber \\ 
&+ V_{\rm T}(\Phi_i,T) + V_{\rm daisy}(\Phi_i,T) \, .
\label{eq:fullpotential}
\end{align}
$V_{\rm tree}$ is the tree-level potential
shown in \cref{eq:HiggsPotential_r2hdm},
$V_{\rm CW}$ consists of zero-temperature radiative
corrections in the form of the usual
Coleman-Weinberg potential renormalized
in the $\overline{\textrm{MS}}$ scheme~\cite{Coleman:1973jx}.
$V_{\rm CT}$ contains UV-finite counterterm contributions
that are defined in such a way as to cancel the first
and second derivatives of $V_{\rm CW}$ with respect
to the fields in the tree-level EW
minimum. This is achieved by expressing $V_{\rm CT}$ in
terms of counterterms for the potential parameters,
\begin{align}
& V_{\rm CT}(\Phi_1,\Phi_2) = \\
& \delta m_{11}^2\,\Pdd + \delta m_{22}^2\,\Puu - \left(\delta m_{12}^2\Pdu + \mathrm{h.c.}\right) \nonumber\\
&\hphantom{=} + \frac{1}{2}\delta\lambda_1 (\Pdd)^2 + \frac{1}{2}\delta\lambda_2 (\Puu)^2  + \delta\lambda_3 (\Pdd)(\Puu) \nonumber        \\
&\hphantom{=} + \delta\lambda_4 (\Pdu)(\Pud) + \frac{1}{2} \left(\delta\lambda_5(\Pdu)^2 + \mathrm{h.c.}\right) \, . \nonumber 
\end{align}
and imposing on-shell-like
renormalization conditions
at zero temperature of the form
\begin{align}
  \label{eq:renoconds1}
\partial_{\phi_{i}}V_{\rm CT}(\Phi)\left|_{\left\langle
\phi\right\rangle_{T=0}}\right.&=-\partial_{\phi_{i}}V_{\rm CW}(\Phi)
\left|_{\left\langle \phi\right\rangle _{T=0}}\right. \ , \\
\partial_{\phi_{i}}\partial_{\phi_{j}}V_{\rm CT}(\Phi)
\left|_{\left\langle \phi\right\rangle _{T=0}}\right.&=
-\partial_{\phi_{i}}\partial_{\phi_{j}}V_{\rm CW}(\Phi)
\left|_{\left\langle \phi\right\rangle _{T=0}}\right. \ ,
  \label{eq:renoconds2}
\end{align}
from which the parameter counterterms can be computed solving
the resulting system of equations
(see Ref.~\cite{Basler:2018cwe} for details).
Here $\phi_{i,j}$ schematically denote the real and imaginary component
fields of the doublets $\Phi_1$ and $\Phi_2$,
and the derivatives are evaluated in the EW minimum at
zero temperature indicated with $\left\langle \phi\right\rangle_{T=0}$.
Imposing these conditions ensures that
the tree-level values
of the vacuum expectation values and the scalar masses
are maintained
at one-loop level~\cite{Basler:2016obg}.
This prescription also facilitates the decoupling of
radiative corrections for heavy 
particles~\cite{Anderson:1991zb}.\footnote{It should
be noted that the true pole masses
of the scalars also receive momentum-dependent corrections
which are neglected if the finite counterterm shifts
are derived from the effective potential.
While the momentum-dependent corrections are
neglected in this approach,
numerically they are expected to be small.}
In contrast to the vevs and the masses,
the trilinear and quartic couplings among the
scalars receive finite corrections at one-loop level
whose cancellation is not enforced by the renormalization conditions
shown in \cref{eq:renoconds1} and \cref{eq:renoconds2}.
Finally, thermal corrections that explicitly
depend on the temperature~$T$ are contained in
$V_{\rm T}$ and $V_{\rm daisy}$~\cite{Quiros:1999jp},
which
include the one-loop thermal corrections and
corrections from the resummation of daisy diagrams,
respectively,
following the Arnold-Espinosa
prescription~\cite{Arnold:1992rz}.

\texttt{BSMPT} traces the temperature-dependent vacuum structure by numerically identifying all
phases, taking into account all four possible
field directions in terms of the vevs shown in
\cref{eq:vevs}. 
If a pair of co-existing vacua in a certain temperature
range is detected, a first-order phase transietion
from the false vacuum $\phi_{i,\mathrm{false}}$ into the true
vacuum $\phi_{i,\mathrm{true}}$ is possible, where $\phi_i$
in the discussion of this section collectively denotes
the scalar background fields.
The critical temperature $T_c$ is defined as the
temperature at which the vacua are degenerate.
The actual phase transition occurs at the lower
nucleation temperature $T_n$, where bubbles of the
true-vacuum phase can nucleate and expand.\footnote{The
differences between nucleation, percolation and completion
temperatures are small for EWPTs in the C2HDM,
where the possible amount of supercooling is very
limited. We therefore focus on the nucleation temperature
in our discussion for simplicity.}

The decay rate per unit volume as a function
of the temperature is given
by~\cite{Coleman:1977py,Callan:1977pt}
\begin{equation}
  \Gamma(T) = A(T) e^{-S_3(T) / T} \, ,
\end{equation}
where $S_3(T)$ is the three-dimensional Euclidean action
of the bounce solution describing the bubble,
and $A(T)$ is a prefactor with mass dimension four.
The condition for the onset of the phase transition is
that on average one bubble nucleates per Hubble volume,
which for EWPTs (approximating the sub-leading
prefactor as $A(T) = T^4$) corresponds 
to~\cite{Caprini:2019egz}
\begin{equation}
  \frac{S_3(T_n)}{T_n} \approx 140 \, .
\label{eq:onset}
\end{equation}
This condition defines
the nucleation temperature $T_n$.
The bounce action $S_3$ is computed by finding the
O(3)-symmetric bounce solution $\phi_{i,b}(\rho)$
that extremises the Euclidean action
\begin{equation}
  S_3(T) = 4 \pi \int_0^\infty d\rho \, \rho^2 \left[
    \frac{1}{2} \left(
      \frac{d \phi_i}{d\rho}
      \right)^2 + V(\phi_i, T)
  \right] \, ,
\label{eq:EuclideanAction}
\end{equation}
where $\rho$ is the radial coordinate of the bubble.
Determining $\phi_{i,b}(\rho)$
amounts to solving the equation of motion
(or bounce equations)
\begin{equation}
\frac{d^2 \phi_i}{d \rho^2} + \frac{2}{\rho} \frac{d \phi_i}{d \rho} = \nabla V(\phi_i) \, ,
\label{eq:BubbleEOM}
\end{equation}
subject to the boundary conditions
\begin{equation}
\phi_i(\rho \to \infty) = \phi_{i,\mathrm{false}}, \qquad \left. \frac{d \phi_i}{d \rho } \right|_{\rho = 0} = 0 \, .
\end{equation}
The code \texttt{BSMPT} solves this set of
differential equations using the path deformation
algorithm~\cite{Wainwright:2011kj}.

The EWPT can be characterized by four key thermal
parameters: the strength of the transition $\alpha$,
the inverse duration in Hubble units $\beta/H$,
the nucleation temperature $T_n$,
and the bubble wall velocity $v_w$~\cite{Auclair:2022lcg}.
The parameter $\alpha$
can be defined
in terms of the trace of the energy-momentum 
tensor, 
\begin{equation}
\alpha = \frac{1}{\rho_R} \left[
  \Delta V(T) - \left.
    \frac{T}{4} \frac{\partial \Delta V(T)}{\partial T}
    \right|_{T_{n}} \right] \, . \label{eqalpha}
\end{equation}
Here, $\Delta V(T) = V_{\rm false} - V_{\rm true}$
with $V_{\rm false} = V(\phi_{i,\mathrm{false}})$
and $V_{\rm true} = V(\phi_{i,\mathrm{true}})$
is the difference in the
effective potential between the false and true vacua,
and \mbox{$\rho_R = \pi g_{\rm eff} T^4 / 30$}
is the background energy density
assuming a radiation dominated universe,
with $g_{\rm eff}$ being the effective degrees of freedom.
Larger values of $\alpha$ correspond to stronger
phase transitions.
The inverse duration $\beta / H$ can be calculated
with
\begin{equation}
\frac{\beta}{H} = T_{n} \left.
  \frac{\partial}{\partial T} \left(
    \frac{S(T)}{T} \right) \right|_{T_{n}} \, , \label{beta}
\end{equation}
where $H$ is the Hubble parameter.

Finally, the bubble wall velocity $v_w$ is a
an important parameter influencing the dynamics of
the phase transition, the resulting BAU, and the
peak frequency and amplitude of the resulting
GW spectrum.
However, unlike the parameters $\alpha$ and
$\beta / H$, it is difficult to compute reliably as
it requires solving out-of-equilibrium transport
equations that account for the friction between
the moving 
wall and the surrounding plasma
(see Refs.~\cite{DeCurtis:2022hlx,DeCurtis:2023hil,
DeCurtis:2024hvh} for recent
progress in the computation of out-of-equilibrium dynamics).
We estimate the bubble wall velocity following the
simplified toy model presented in
Ref.~\cite{Dorsch:2016nrg}.
In particular, we use the relation between $v_w$
and the strength parameter $\alpha$ derived in that
work for the C2HDM in order to assign a
value of $v_w$ to each parameter point featuring
an EWPT. Typical values for the wall velocity in our
setup fall within the range
$0.1 \lesssim v_w \lesssim 0.25$.
This interval agrees fairly well with
the values for $v_w$ that
were found for a selection of parameter points
the inert 2HDM in Ref.~\cite{Jiang:2022btc,
Ekstedt:2024fyq}, whereas larger values of
$v_w \approx 0.6$ were predicted
for the inert 2HDM in
Ref.~\cite{Branchina:2025jou}.

\subsubsection{Electroweak baryogenesis}
\label{sec:ewbg}

The baryon asymmetry can be dynamically generated during a strong first-order EWPT via EWBG~\cite{Kuzmin:1985mm}.
An essential condition for successful EWBG is a
sufficiently strong EWPT. This is typically
quantified in terms of the baryon number preservation
criterion (or baryon washout 
condition)~\cite{Dimopoulos:1978kv,Kuzmin:1985mm}
\begin{equation}
\xi_n \equiv \frac{v_n}{T_n} \gtrsim 1 \, ,
\label{eq:bnpc}
\end{equation}
where, in the C2HDM, the EW symmetry breaking vev in
the true minimum at the nucleation temperature $T_n$
is given by
\begin{equation}
v_n = v(T_n) = \sqrt{v_1^2(T_n) + v_2^2(T_n) +
v_{\rm CP}^2(T_n)} \, .
\end{equation}
The washout condition ensures that baryon-number violating
sphaleron processes are suppressed in the
broken phase and that the generated BAU is preserved
after the transition completes.\footnote{The
baryon number preservation criterion defined
in this way is known to exhibit a residual gauge
dependence~\cite{Patel:2011th}.
In this work, we evaluate the criterion
using the effective potential in Landau gauge,
where the impact of this dependence is expected to be
subdominant compared
to other sources of theory uncertainty~\cite{Garny:2012cg},
such as renormalization scale dependence,
missing higher-order corrections, and ambiguities in
the thermal resummation procedure~\cite{Bittar:2025lcr}.}

Instead of solving the full set of transport equations
to compute the BAU in the C2HDM,
we make use of the results
from Refs.~\cite{Fromme:2006cm,Fromme:2006wx,Dorsch:2016nrg}
in which the transport
equations were studied with the CP-violating
source term computed in the WKB approximation~\cite{Joyce:1994fu, Joyce:1994zt}
and focusing on the top-quark transport which
is the relevant contribution in the C2HDM.
To estimate the baryon asymmetry generated during the EWPT, we adopt the analytic approximation introduced in Ref.~\cite{Huber:2022ndk}.
In this framework, the baryon-to-entropy ratio is given by 
\begin{equation} 
\frac{\eta_s}{10^{-11}} \approx 6 \cdot 10^2 \, \frac{\sin(\delta_t)  \xi_n^2}{L_w T_n} \, ,
\label{eq:eta_estimate}
\end{equation}
where $L_w$ is the thickness of the bubble wall and $T_n$ is the nucleation
temperature.\footnote{$\eta_s$ is
related to $\eta_\gamma$, see \cref{eq:bau}, via
the relation $\eta_s \approx \eta_\gamma / 7.04$.}
The phase $\delta_t$ encodes the CP-violating source relevant for baryogenesis, and depends on $t_\beta$ through the top-quark Yukawa coupling,
\begin{equation} 
\delta_t = \frac{\Delta \theta}{1 + t^2_\beta} \, ,
\label{eq:deltat}
\end{equation}
with $\Delta \theta$ denoting the variation of the CP-violating phase of the top Yukawa across the bubble wall~\cite{Fromme:2006cm} \begin{equation} \Delta \theta = \theta_{\rm true} - \theta_{\rm false} . \end{equation}

Working in the basis in which the CP-violating vev
is contained in $\Phi_2$, see \cref{eq:vevs},
the CP phase in the true minimum can be computed
with the vevs in the global minimum, such that
\begin{equation}
\theta_{\rm true} = \tan^{-1} \left(
  \frac{v_{\rm CP, true}}{v_{2,{\rm true}}}
  \right) \, .
\end{equation}
For the CP phase in the false minimum, this formula
cannot be applied because in the C2HDM we find
(in most cases) $v_{2, {\rm false}} = 0$.
Instead, we compute $\theta_{\rm false}$
using the field profile of the bounce solution
$\phi_i(\rho)$, see the discussion
in \cref{sec:ThermalHistory},
in the limit $\rho \to \infty$, i.e~\cite{Huber:2022ndk}
\begin{equation}
  \lim_{\rho \to \infty} = \tan^{-1} \left(
    \frac{\phi_{\rm CP}(\rho)}{\phi_{2}(\rho)}
    \right) \, ,
\end{equation}
where the field profiles $\phi_{\rm CP}(\rho)$ and
$\phi_2(\rho)$ where obtained by
solving the bounce equations
as discussed in \cref{sec:ThermalHistory}.
Finally, we compute $L_w$
following Ref.~\cite{Dorsch:2016nrg}
by fitting the field profile of $\phi_2(\rho)$
with a kink profile of the form
\begin{equation}
 \phi_2(\rho) = \frac{\phi_{2}(0) - 
    \phi_{2}(\infty)}{2}
  \left(
    1 - \tanh \frac{\rho - \rho_0}{L_w}
  \right) + \phi_{2}(\infty) \, ,
\end{equation}
with $\rho_0$ being and $L_w$ being
the parameters that are fitted.
Since the true bubble profile is only
approximated by the kink profile,
as a cross check, we fitted also $\phi_1(\rho)$ with
the same kink profile to verify if the results
for $L_w$ are consistent. The values for $L_w$ obtained
with either $\phi_2(\rho)$ or $\phi_1(\rho)$ were
in good agreement.

The results of Refs.~\cite{Fromme:2006wx,Dorsch:2016nrg},
and consequently also the
estimate of the baryon-to-entropy ratio~$\eta_s$
shown in \cref{eq:eta_estimate}, rely on the gradient
expansion, whose validity breaks down for too thin
bubbles with $L_w T_n \lesssim 2$. We find
for the EWPT in our benchmark planes values
of $L_w T_n > 2$, see the discussion in
\cref{app:Lw}, such that the gradient expansion
is applicable. Nevertheless, we stress that
the approximation applied here
provides only a qualitative estimate of the
BAU, enabling us to identify regions of parameter
space that are favoured by a successful realisation
of EWBG. Should future LHC data provide indications favouring a specific realisation of the C2HDM that allows a sufficiently strong EWPT, it would be well motivated to pursue a more refined and quantitatively
more robust computation of the baryon asymmetry within that particular scenario.

While our procedure to compute $L_w$
provides an estimate of the wall
width computed at nucleation,
we stress that it does not capture the
subsequent stretching of the wall during bubble
expansion due to plasma friction and 
pressure~\cite{Moore:1995si}. 
Previous studies have shown that, for not too
large bubble wall velocities, such
bubble expansion effects only lead to small
changes in the bubble profile~\cite{Konstandin:2013caa}.
Specific models have been analysed
in Ref.~\cite{Konstandin:2014zta}, and the
increase of the values of $L_w T_n$ compared
to the ones obtained at nucleation
is at the level of 10 to 50\%.
As will be further discussed in \cref{sec:results},
to account for this and other uncertainties
in the predictions for the BAU,
we indicate parameter regions
in our plots in which the predicted BAU is
within factors of 0.5 and 2.0 in agreement
with the measured value.
We have checked explicitly  that
even if $L_w T_n$ were underestimated by a
factor of 2, the BAU-favoured regions of
parameter space would only be slightly reduced,
not affecting our conclusions.
It should also be noted that, since the plasma
friction increases the bubble wall width,
the gradient expansion that underlies the
theory prediction for the BAU remains valid
if $L_w T_n \gtrsim 2$ is satisfied with
$L_w$ calculated at nucleation.

An additional source of uncertainty on the prediction for
the BAU arises from its dependence on the
bubble wall velocity $v_w$.
For highly relativistic bubble wall
velocities, $v_w \approx 1$, the generated BAU is
generally suppressed~\cite{Fromme:2006wx}, which weakens
the prospects for a combined realization of EW
baryogenesis and detectable GW signals.
However, there is still some debate about whether
a phase transition with supersonic bubble walls,
$v_w \gtrsim c_s \approx 0.6$, where $c_s$
is the speed of sound in the relativistic plasma,
could still generate a sufficient amount of
BAU~\cite{Cline:2020jre}.
In this context,
Ref.~\cite{Dorsch:2021ubz}
added additional support for this claim,
suggesting that using a generalized fluid Ansatz
and taking into account higher moments in the
treatment of the Boltzman equations
makes this scenario more
viable, whereas Ref.~\cite{Kainulainen:2024qpm}
argues that the BAU is more strongly suppressed
when higher-order moments, beyond the standard
two-moment expansion in the transport equations,
are included.
In our analysis,
we estimate the baryon asymmetry using an
approximation valid in the regime of
slow-moving (sub-sonic) bubble walls
where there is only a minor dependence
on $v_w$~\cite{Fromme:2006cm,Dorsch:2016nrg}.
Using the method discussed in \cref{sec:ThermalHistory},
the predicted bubble wall velocities remain
below $v_w \approx 0.3$,
well within the range where the approximation
remains reliable and only mildly depends on~$v_w$.
Therefore, we do not consider an additional
suppression from fast-moving bubble walls.

\subsubsection{Gravitational wave relics}

First-order cosmological phase transitions generate a
stochastic background of GWs~\cite{Witten:1984rs, Hogan:1986qda},
potentially observable by future space-based GW
interferometers. In the case of an EWPT, the resulting
GW spectrum typically peaks in the millihertz range,
precisely where detectors like LISA are most
sensitive~\cite{LISA:2017pwj, LISACosmologyWorkingGroup:2022jok}.

In the 2HDM, the dominant contributions to the GW signal arise not from the bubble wall collisions themselves, but from the bulk motion of the thermal plasma induced by expanding bubbles. These include long-lasting sound waves and magnetohydrodynamic turbulence~\cite{Dorsch:2016nrg}. To model the resulting GW spectrum, we use numerical power-law fits derived from hydrodynamical simulations, expressing the spectral shape, amplitude, and peak frequency as functions of the four key phase transition parameters discussed in \cref{sec:ThermalHistory}.
The specific formulas for the spectral shape
of the total GW signal $h^2 \Omega_{\rm GW}$
are taken from Ref.~\cite{Biekotter:2022kgf},
which 
uses the results presented
in Refs.~\cite{Caprini:2019egz, Auclair:2022jod}.
We consider the contributions from sound waves
and turbulence,
\begin{equation}
    h^2 \Omega_{\rm GW} = h^2 \Omega_{\rm sw} + h^2 \Omega_{\rm turb} \, ,
\end{equation}
where the dominant contribution is $h^2 \Omega_{\rm sw}$.

The detectability of 
a stochastic GW signal is determined by the
signal-to-noise ratio (SNR) given by
\begin{equation}
{\rm SNR} = \sqrt{\mathcal{T} \int_{- \infty}^{+ \infty} {\rm d}f \left[\frac{h^2\Omega_{\rm GW}(f)}{h^2\Omega_{\rm Sens}(f)}\right]^2} \, ,
\end{equation}
where $\mathcal{T}$ is the exposure time, and
$h^2 \Omega_{\rm Sens}$ denotes the
nominal detector sensitivity
to stochastic sources~\cite{LISAmissionreq}.
In our analysis, we focus on LISA and assume a nominal mission duration of $\mathcal{T} = 7$ years.
A stochastic GW typically is considered to be detectable
if the SNR is larger than one.
However,
taking into account the substantial theoretical
uncertainties in the predictions for the GW 
signals~\cite{Croon:2020cgk,Athron:2022jyi,Lewicki:2024xan},
we consider a GW signal to be potentially detectable
with LISA if the SNR exceeds the
threshold ${\rm SNR} > 10^{-3}$, as further discussed below.

\section{Results}
\label{sec:results}

In this section, we present our numerical results
for the benchmark planes that we define in the C2HDM.
We explore parameter space regions where a strong
first-order EWPT is realised.
Among these, we indicate where the BAU
can be explained via EWBG, and we indicate where the
GW signal is predicted to be sufficiently
strong to be potentially detectable with LISA.

In our analysis,
we only considered the final phase transition
present in the thermal history.
While it is possible to predict multiple phase transitions
in the 2HDM~\cite{Lee:2025hgb,Bhatnagar:2025jhh},
earlier transitions tend to be either non-first-order or
have a weaker strength compared to the final one.
Combined with the fact that these earlier transitions
happen at a higher temperature,
their contribution to the GW signal can be expected to
be significantly smaller than the one of the final
phase transition. Regarding EW baryogenesis,
we verified that the EWPT
starts from an EW-symmetric vacuum, with all vevs
equal to zero. Otherwise sphaleron processes are
potentially inactive
and no BAU could be produced during the transition.
Accordingly, even if there were
previous strong first-order EWPTs in the thermal history,
they do not have an impact on the final
prediction for the BAU as any asymmetry would be washed
out prior to the final EWPT.
Therefore, our approach to focus only on the final phase
transition does not overlook any relevant contributions to the
GW signal or the BAU.

For each parameter point that yields a first-order EWPT with $\xi_n > 0.8$, the corresponding value of $\xi_n$ is represented by the colour of the point.
We choose a threshold value of 0.8,
which is slightly smaller than the value 1
appearing in the baryon number
preservation criterion shown in \cref{eq:bnpc}, since
the predictions for $\xi_n$ are subject to sizable
theoretical uncertainties. In particular, while
we employ the Arnold-Espinosa resummation prescription,
it was shown recently for the real 2HDM
in Ref.~\cite{Bittar:2025lcr} that
a more consistent thermal
resummation via partial dressing predicts smaller
nucleation temperatures, and therefore larger values
of $\xi_n$ compared to the Arnold-Espinosa prescription.

Similarly, the predictions for the baryon-to-entropy
ratio $\eta_s$ are subject to large uncertainties,
not only from those related to the
parameters characterising the EWPT, such as
as the bubble wall velocity
and the bubble wall
width~\cite{Fromme:2006wx,Cline:2020jre,
Dorsch:2021ubz},
but also from uncertainties associated with the derivation of the
CP-violating source terms in the transport
equations~\cite{Postma:2022dbr}.
Taking this into account, the blue lines in our plots indicate
the parameter ranges that predict strong
first order EWPT and values of $\eta_s$ in the range of
range $\eta_s^{\rm exp} / 2 \leq \eta_s \leq 2
\eta_s^{\rm exp}$, with $\eta_s^{\rm exp} \approx
8.7 \cdot 10^{-11}$, which should be seen as
as regions indicating a successful
of the EWBG.

Regarding GWs, we adopt a threshold of
\mbox{$\mathrm{SNR}>10^{-3}$} for identifying potentially detectable
GW signals with LISA, rather than the commonly used
threshold of $\mathrm{SNR}>1$. This choice is motivated by
significant theoretical uncertainties in the prediction
of the GW spectrum. For instance, our computation of
the bubble wall velocity $v_w$ yields subsonic values
$v_w \lesssim 0.2$ for the bulk of parameter points,
based on the approach detailed in \cref{sec:ThermalHistory}.
However, the dominant sound wave contribution
to the GW signal peaks for larger
velocities, $v_w \approx c_s \approx 0.6$.
If the true bubble wall velocity is underestimated
in our setup,
the predicted SNRs could be smaller
by one to two orders of magnitude compared to scenarios
where $v_w \approx c_s$ is
assumed~\cite{Biekotter:2022kgf}.\footnote{Recent
numerical real-time
simulations of the bubble growth beyond the
local thermal equilibrium approximation show
detonation with $v_w \approx 1$ instead of
deflagration solution $v_w \lesssim c_s$
under certain conditions~\cite{Krajewski:2024zxg}.}
Moreover, recent work~\cite{Bittar:2025lcr} in the R2HDM
using a more refined thermal resummation scheme called partial dressing~\cite{Curtin:2016urg}
has found stronger EWPTs than those
predicted with the Arnold–Espinosa method employed
here. Although the parameter regions favourable for
an EWPT remain unchanged, the nucleation temperatures
using the partial dressing approach were found to
be lower by a factor of two in the considered
benchmark point. This would enhance the GW signal
by roughly two orders of magnitude. 
In light of these considerations we use
a lower SNR threshold to identify scenarios that
may yield observable signals at LISA once theoretical
uncertainties are better understood.\footnote{The
detectability of a predicted GW signal ultimately depends
on additional theoretical uncertainties from the
hydrodynamics of GW production and experimental uncertainties
such as unknown astrophysical or cosmological foregrounds.
These considerations lie beyond the scope of this work.}

\subsubsection{Smoking gun scenario}
\label{sec:bp-sg}

The first benchmark plane that we define is
motivated by results obtained in the
CP-conserving R2HDM. In the R2HDM, the strongest
EWPT were found for scenarios featuring the
alignment limit
$\cos(\beta - \alpha) \approx 0$ and a
sizable mass gap with
$m_A > m_H$, with $m_{H^\pm} \approx m_A$,
and $M \approx m_H$ to satisfy theoretical
and experimental constraints, see the discussion
above.
We construct our C2HDM benchmark plane by
fixing the scalar
masses based on these R2HDM results,
and we study the impact
of CP violation in the C2HDM by varying
the mixing angles $\alpha_1$ and $\alpha_3$.
Specifically, we vary $\alpha_1$ in a range around
the alignment condition, $\sin(\beta-\alpha_1) \approx 0$.
The other mixing angle $\alpha_3$
used as input parameter is varied
around $\alpha_3 \approx 0$.
Then, the lighter BSM neutral scalar
$h_2$ remains predominantly CP-even (corresponding
to the state $H$ in the R2HDM limit), while the
heavier neutral BSM scalar $h_3$ remains
predominantly CP-odd (corresponding to the state
$A$ in the CP-conserving R2HDM).

\begin{figure*}
\subfloat[
  \textbf{BP-SG}: smoking-gun scenario for
  a strong first-order EWPT with a mass gap between
  a lighter and mainly CP-even state~$h_2$
  and a heavier mainly CP-odd state~$h_3$.]
  {
    \label{fig:smoke_normal}
    \includegraphics[width=0.62\textwidth]{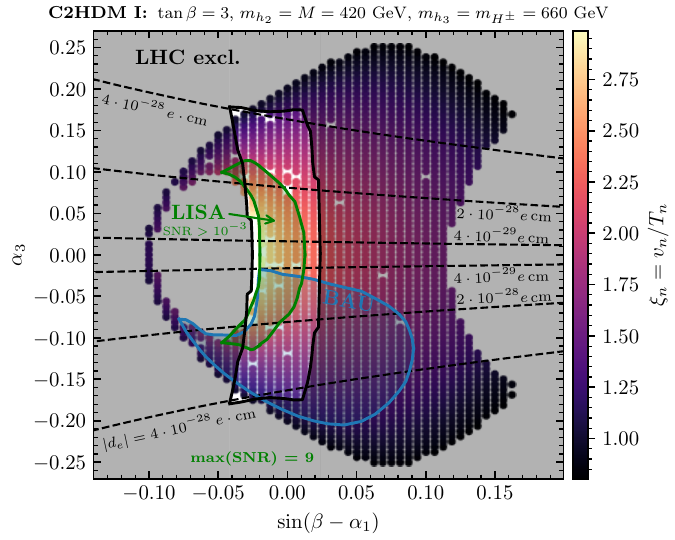}
  } \\[1em]
\subfloat[
  \textbf{BP-ISG}: inverted smoking gun scenario
  with a ligher and mainly CP-odd state~$h_2$
  and a heavier mainly CP-even state~$h_3$.]
  {
    \label{fig:smoke_inverted}
    \includegraphics[width=0.62\textwidth]{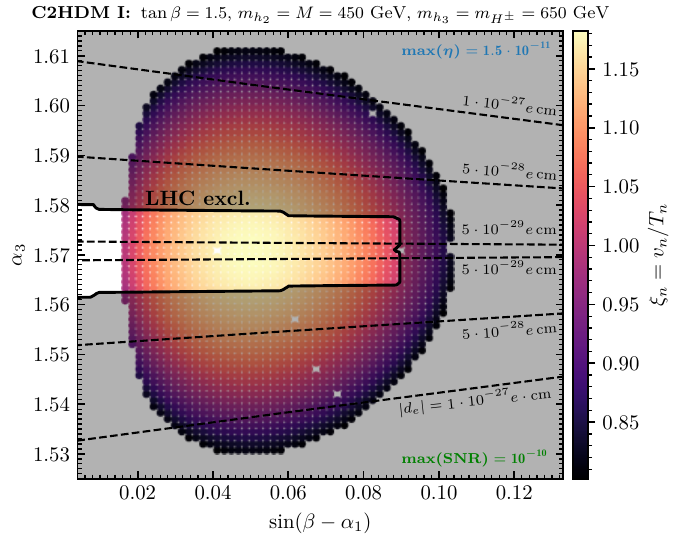}
}
\caption{
Shown are parameter points in the $\{\sin(\beta - \alpha_1), \alpha_3\}$ 
planes that predict a strong first-order EWPT, with the 
color of each point indicating the corresponding value of $\xi_n$. 
The remaining C2HDM parameters are fixed as specified in the plot headers.
The grey shaded areas are excluded by cross section
limits from BSM scalar searches at the LHC.
The green area indicates parameter space regions
featuring a GW signal with a LISA SNR larger than
SNR~$>10^{-3}$ assuming an exposure time of
7 years. Taking into account the sizable theoretical
uncertainties, these areas can potentially
be probed with LISA
(see the text for details). The blue area indicates
parameter regions
with the baryon-to-entropy
ratio $\eta_s$ predicted to
be in the range $\eta^{\rm exp}_s / 2 \leq \eta_s \leq
2\eta^{\rm exp}_s$. Finally, the dashed black lines
are contour lines indicating the prediction for
the electron EDM $|d_e|$. In the lower plot no green and blue
regions are visible because the EWPT is not sufficiently
strong to give rise to a detectable GW signal or to
a baryon-to-entropy ratio in agreement with observations,
respectively.}
\label{fig:cpsmoker1}
\end{figure*}

Concretely, the benchmark plane smoking gun 
(\textbf{BP-SG})
is defined by
\begin{align}
\textrm{\textbf{BP-SG:} }&-0.3 \leq s_{\beta - \alpha_1} \leq 0.3 \, , \;
-0.2 \leq \alpha_3 \leq 0.2 \, , \notag \\
&t_\beta = 3, \; m_{h_{2}} = M = 420\GeV \, , \notag \\
&m_{h_{3}} = m_{H^\pm} = 660\GeV \, .
\end{align}
The size of the splitting between $m_{h_2}$
and $m_{h_3}$ is chosen to be in the interval
where we found a strong EWPT in the R2HDM
in Ref.~\cite{Biekotter:2023eil}, associated
with a GW signal potentially in reach of LISA.
The value of
$t_\beta = 3$ is sufficiently large to avoid
the current cross-section limits from the
smoking gun signature $h_3 \to Z h_2$ with
$h_2 \to t \bar t$ performed by ATLAS and
CMS~\cite{ATLAS:2023zkt,CMS:2024yiy},
but small enough to not suppress
significantly the predicted BAU, which
approximately scales
with a factor of $1 / t_\beta^2$ as shown in
\cref{eq:deltat}.

The results for the benchmark plane
\textbf{BP-SG} are shown
in \cref{fig:smoke_normal}.
We find an area featuring a strong EWPT
for values of $-0.10 \lesssim s_{\beta - \alpha_1}
\lesssim 0.16$ and $|\alpha_3| \lesssim 0.25$.
Close to the CP-conserving limit $\alpha_3 \approx 0$,
this region only spans to smaller values of
$s_{\beta - \alpha_1} \approx 0.11$ because there the
one-loop potential becomes unbounded and the
EW vacuum at zero temperature is unstable.\footnote{The
impact of vacuum instabilities would be more severe
considering only the tree-level potential, excluding parameter
points with an EWPT for values of 
$s_{\beta - \alpha_1} \gtrsim 0.025$
at $\alpha_3 \approx 0$ and values of
$s_{\beta - \alpha_1} \gtrsim 0.090$
at $\alpha_3 \approx \pm 0.25$
(see the discussion in \cref{sec:theoconstraints}).}
Larger values of $\xi_n$ correspond to stronger EWPTs.
Within the region of parameter space predicting a first-order EWPT, the values of $\xi_n$ (defined in \cref{eq:bnpc} and indicated by the colour scale) tend to increase as $|s_{\beta - \alpha_1}|$ and $\alpha_3$ decrease, reaching their maximum near the alignment limit where the Higgs boson $h_1$ has SM-like couplings.
This can be understood intuitively by the fact 
that the scalars involved as dynamical fields in the 
EWPT should be as light as possible in order 
to increase the strength of the phase transition~\cite{Dorsch:2017nza}, 
just as in the SM, where the strengths 
decrease with increasing Higgs boson mass 
(with all other parameters fixed)~\cite{Kajantie:1996mn}.
In the alignment limit, the mass eigenstate basis
is identical to the Higgs basis~\cite{Gunion:2002zf},
in which only the lightest scalar $h_1$ obtains a vev
during EW symmetry breaking. However, for
$\beta \neq \alpha_1$ and/or $\alpha_3 \neq 0$,
also the field directions corresponding to the
heavier states $h_2$ and/or $h_3$ would obtain
a vev, respectively, which weakens the
strength of the EWPT.

In the area of points with an EWPT, we find
a nested region in the interval
$-0.10 \lesssim s_{\beta - \alpha_1} \lesssim -0.03$
where no EWPT is predicted,
and accordingly no points are displayed.
In this nested area the condition shown in
\cref{eq:onset} for the on-set of the phase transition
is never met due to a too large potential barrier
separating the true and false minima, and
the universe remains trapped in the unphysical
vacuum with unbroken EW symmetry although a deeper
EW-symmetry breaking minimum exists at
$T = 0$. This \textit{vacuum trapping} effect has been
studied in detail in the R2HDM in
Ref.~\cite{Biekotter:2022kgf},
highlighting that it excludes
parameter space regions in which otherwise the
potentially strongest EWPTs would be found, thus
diminishing the prospects for the detection of
GWs at LISA.\footnote{When thermal
fluctuations are insufficient to trigger the
EWPT, it could in principle still occur via
quantum tunneling~\cite{Coleman:1977py}.
However, if the potential
barrier suppresses thermal transitions, the vacuum
decay rate from tunneling is similarly suppressed
and cannot initiate the transition.~\cite{Coleman:1977py}.
Moreover, after significant
amount of supercooling,
if the plasma temperature drops below
the Hubble scale,
the tunneling rate should be evaluated
taking into account de Sitter curvature
effects~\cite{Kearney:2015vba,Joti:2017fwe}.
However, for an EW-scale vacuum, the Hubble
parameter is too small to compete with the
barrier height, yielding exponentially suppressed 
rates. An onset of the EWPT via de Sitter
fluctuations would therefore occur only after
vacuum domination, preventing successful percolation. 
This is known as the \textit{graceful-exit problem}
of old inflation~\cite{Guth:1982pn}.
An EWPT triggered by de Sitter fluctuations
is therefore not phenomenologically
viable (see also the discussions
in Refs.~\cite{Cline:1999wi,
Ellis:2018mja,Biekotter:2021ysx}).}
We observe here in the C2HDM that for all values of
$s_{\beta - \alpha_1}$ and $\alpha_3 \approx 0$
where the vacuum trapping
forbids the presence of an EWPT,
there is a non-vanishing value of $\alpha_3$
where the EWPT takes place. 
This indicates that the C2HDM is able to mitigate
the impact of vacuum trapping observed in the
R2HDM by introducing small CP-violating mixings
via $\alpha_3 \neq 0$.

Substantial parts of the parameter region
predicting a strong EWPT are already excluded
by LHC constraints, whose exclusions are indicated
with the grey shaded area. Only the 
area within the black solid lines passes the
limits from all BSM scalar searches
included in \texttt{HiggsTools}.
The parameter regions to the left and to the right
of the allowed area are excluded by CMS searches
for the resonant pair production of the
125~GeV Higgs boson via the decay
$h_2 \to h_1 h_1$,
based on a CMS combination of searches
in $b \bar b b \bar b$, $b \bar b \tau^+ \tau^-$,
$b \bar b \gamma \gamma$ and $b \bar b W^+ W^-$
final states~\cite{CMS:2024phk}.
The regions above and below the allowed region
are excluded by searches for $h_2 \to Zh_1$ with
subsequent decay $h_1 \to b \bar b$,
performed by ATLAS~\cite{ATLAS:2022enb}.
Within the allowed region, the currently most
sensitive LHC searches are the ones for the
smoking gun signature $h_3 \to Z h_2$ with the
lighter BSM state $h_2$ decaying into $t \bar t$
pairs~\cite{ATLAS:2023zkt,CMS:2024yiy}, for which
the cross sections remain sizable in and around the 
alignment limit because only
BSM scalars are involved.
The fact that searches for multi-Higgs signatures 
(where either the SM-like Higgs or another scalar is produced 
in the decay of a heavier BSM spin-0 resonance) already 
exclude significant portions of the otherwise viable parameter space, 
or constitute the most sensitive probes of the remaining regions, 
highlights their increasing relevance for testing the 
electroweak phase transition in future LHC runs.
We also find them in this benchmark plane to be significantly 
more constraining
than the limits from the cross-section measurements
of the 125~GeV Higgs boson, which also become relevant
away from the alignment limit.

Within the region consistent with current LHC
constraints, we find that successful EWBG can be
achieved, as indicated by the blue solid line
that encloses the parameter points predicting
$\eta^{\rm exp}_s / 2 \leq \eta_s \leq
2\eta^{\rm exp}_s$.
To the right and bottom of this
blue region, 
the observed BAU cannot be reproduced because the 
EWPT becomes too weak, leading to smaller values of $\xi_n$, 
which enters quadratically in the numerator of \cref{eq:eta_estimate}, 
and larger values of $L_w T_n$ in the denominator. 
To the left of the blue region, 
vacuum trapping prevents the transition from completing, 
therefore precluding 
EWBG. Finally,
in the parameter region on top of
the blue region,
where $|\alpha_3|$ is very small, the
amount of CP violation (contained in the
parameter $\delta_t$ in \cref{eq:eta_estimate})
is not sufficient to facilitate EWBG.

The parameter region suitable for EWBG is
associated with detectable
values of the electron EDM, indicated
in \cref{fig:smoke_normal} with
the black dashed contour lines.
We find that in the blue region the
electron EDM values
exceed $|d_e| = 4 \cdot 10^{-29} e
\cdot \text{cm}$, in tension with the currently
strongest experimental bound of $|d_e| < 4.1 \times
10^{-30} \, e \cdot\text{cm}$~\cite{Roussy:2022cmp}
by about an order of magnitude or more.
Hence, in the absence of cancellation mechanisms
beyond the C2HDM,
the electron EDM constraint excludes the parameter 
region favourable for EWBG in the \textbf{BP-SG} plane.
Alternatively, new sources of CP violation can
be considered that facilitate EWBG without
inducing values of
$|d_e|$ above the experimental limits, in which
case in this plane the parameter space region
with vanishing $\alpha_3$ would become suitable
for EWBG.
(see the discussion in \cref{sec:edm} for more
details).

We also indicate in \cref{fig:smoke_normal} with
the green solid line the
parameter space regions in which the EWPT
is sufficiently strong to produce a primordial
GW background that would potentially be detectable
at LISA, with a $\textrm{SNR} > 10^{-3}$ as
discussed above.
The SNRs of the predicted GW signals peak around
the alignment limit, with a maximum value of
$\text{SNR}_{\max} \approx 9$.
We find an overlap between the blue region
and the green region that is compatible with the
current LHC constraints from BSM scalar searches
and the cross section measurements of the
125~GeV Higgs boson.
This demonstrates that although the presence of
CP violation required for EWBG reduces the strength
of the EWPT in this scenario, the EWPT is still
strong enough to produce a potentially detectable
GW signal. We note here that using a stronger
requirement of $\textrm{SNR} > 1$ would not change
this conclusion. However, the overlap between the
green and the blue areas would shrink substantially.
The overlapping blue and green region is already
partially ruled out by LHC searches for resonant Higgs boson pair production, see the discussion above. Moreover, it is expected to be within reach of upcoming LHC Run~3 searches for the smoking gun
process $h_3 \rightarrow h_2 Z \rightarrow (t\bar{t})  Z$~\cite{Biekotter:2023eil,Arco:2025ydq}.
These searches can be evaded if the value of
$t_\beta$ is increased, but this would also lower
the predicted BAU. In \cref{sec:bp-asg} we discuss
the $t_\beta$-dependence of the predicted BAU
in more detail.

\subsubsection{Inverted smoking gun scenario}
\label{sec:bp-isg}

Now we turn to the analysis of a scenario that
we refer to as the benchmark plane with an
inverted smoking gun signal~(\textbf{BP-ISG}).
In contrast to the typical smoking gun scenario
discussed above, we consider mixing angles such
that the heaviest neutral scalar is dominantly
CP-even, while the lighter BSM neutral scalar is
mainly CP-odd, effectively inverting the would-be 
CP
assignments compared to the case in \textbf{BP-SG}.
This configuration was found to be less
prone to realise a strong EWPT in the
CP-conserving R2HDM~\cite{Dorsch:2014qja}.
The main difficulty arises because, for the same
value of $t_\beta$, the mass gap between the
two BSM scalars $h_2$ and $h_3$ cannot be made as
large without violating theoretical constraints from
perturbative unitarity, a problem that becomes
increasingly severe the larger $t_\beta$.
Moreover, another complication observed in the
R2HDM is that away from the alignment limit,
in \textbf{BP-ISG}
the heaviest state~$h_3$ acquires a vev, which tends to
induce a greater suppression on the strength of the EWPT compared to
the typical smoking gun setup \textbf{BP-SG},
where the second-lightest state $h_2$ obtains a vev.
In this study, we investigate whether the inverted
smoking gun scenario can still support a sufficiently
strong EWPT to account for the observed BAU and/or
produce GW signals detectable by
LISA. As before, we also analyse the impact of CP
violation on the strength of the EWPT, 
going beyond previous studies in
the R2HDM.

In \cref{fig:smoke_inverted}, we show the
results of a benchmark scenario of the 
inverted smoking gun, defined by the parameter plane
\begin{align}
\textrm{\textbf{BP-ISG:} }&0 \leq s_{\beta - \alpha_1} \leq 0.15 \, , \notag \\
&\pi/2 - 0.05 \leq \alpha_3 \leq \pi/2 + -0.05 \, , \notag \\
&t_\beta = 1.5, \; m_{h_{2}} = M = 450\GeV \, , \notag \\
&m_{h_{3}} = m_{H^\pm} = 650\GeV \, .
\end{align}
Due to the values of $\alpha \approx \pi / 2$,
the CP assignments of the neutral BSM
Higgs bosons are flipped compared to the scenario
discussed in \cref{sec:bp-sg}: here $h_2$ is
predominantly CP-odd and lighter, while $h_3$ is
heavier and mostly CP-even. As a result, the process
$h_3 \rightarrow h_2 \, Z$ resembles the decay
$H \rightarrow A \, Z$ in the CP-conserving limit.

As seen in \cref{fig:smoke_inverted},
the strongest first-order EWPTs occur in the vicinity
of $\alpha_3 \approx \pi/2$,
where the CP mixing is minimal, 
and the mass eigenstates approximately align
with the CP eigenstates.
Comparing \cref{fig:smoke_inverted} with
\cref{fig:smoke_normal}, one can observe that
the interval of $1.53 \lesssim \alpha_3 \lesssim 1.61$
that can predict a strong EWPT
is significantly smaller than the $\alpha_3$-range
compatible with a strong EWPT in the smoking-gun scenario.
This indicates that the inverted smoking-gun scenario
shows less compatibility between the presence of an
EWPT and CP-violation in the Higgs potential.

Overall, the strongest EWPTs are centered around  the
alignment limit $s_{\beta - \alpha_1} \approx 0$
and $\alpha_3 \approx \pi / 2$, occurring at slightly
positive values of $s_{\beta - \alpha_1}$.
However, this shift away from  $s_{\beta - \alpha_1} = 0$
is so small that the modifications
of the couplings of the 125~GeV Higgs are
well in agreement with LHC cross section 
measurements~\cite{CMS:2022dwd,ATLAS:2022vkf}.
For instance, modifications of the coupling of $h_1$ to
massive gauge bosons remain below 1\%, and the couplings
of $h_1$ to fermions shows modifications of at most
5\% in the parameter space region not excluded by
LHC searches for additional Higgs bosons (see the
discussion below).
We thus conclude that the \textbf{BP-ISG} supports
the general expectation in the C2HDM that a strong
EWPT is associated with a 125~GeV Higgs boson
whose properties, within current LHC experimental precision,
remain indistinguishable from those of a SM Higgs boson.

The values of $\xi_n$ that we find in this benchmark
plane are at most reaching $\xi_n \approx 1.2$,
thus barely satisfying the condition shown in
\cref{eq:bnpc} that prevents the washout of the
BAU after the transition, and which
is required for a realisation of EWBG.
As a consequence of the smaller strengths of
the EWPTs and the smaller amount of CP-violation
due to the more restricted range for $\alpha_3$,
no region in the benchmark plane reaches a BAU
sufficient to match the observed baryon-to-entropy ratio,
as indicated by the absence of blue contours
in \cref{fig:smoke_inverted}.
Additionally, there is no green area because
the GW signals are 
predicted to be much weaker across the parameter space,
with a maximum LISA SNR of $\text{SNR} \approx 10^{-10}$,
far below the sensitivity threshold of LISA.
This is consistent with the relatively weak strength of
the phase transition, as encoded in the moderate
values of $\xi_n \approx 1$.

In principle, the prospects for a strong EWPT in
this inverted smoking-gun scenario could be improved by
increasing the mass splitting between $h_2$ and $h_3$,
which tends to enhance the strength of the transition.
This can be achieved by simultaneously choosing smaller
values of $t_\beta$, thereby avoiding a strongly coupled 
regime and maintaining agreement with perturbative
unitarity. However, reducing $t_\beta$ also intensifies
the tension with current LHC searches,
while the absence of signals for BSM Higgs bosons already
imposes stringent constraints on this region
of parameter space.

The parameter regions that are excluded by the current
cross section limits from LHC searches are shaded in grey,
and only the small region within the solid black
line is still allowed.
The shape of the allowed parameter region in this
scenario is largely dictated by an ATLAS combination of
searches targeting the resonant pair production of the
125~GeV Higgs boson in the $b\bar{b}b\bar{b}$,
$b\bar{b}\tau^+\tau^-$, and $b\bar{b}\gamma\gamma$
final states~\cite{ATLAS:2023vdy}, which exclude the areas
above and below the allowed band.
In addition, ATLAS searches for spin-0 resonances
decaying into 
a $Z$ boson and a 125~GeV Higgs boson~\cite{ATLAS:2022enb}
exclude points with a strong EWPT located to the right
of this region.
A key difference compared to the previously discussed 
benchmark plane in \cref{sec:bp-sg} lies in the nature
of the decaying state $h_2$.
In the inverted smoking-gun scenario
\textbf{BP-ISG} discussed here,
the relevant decay is $h_2 \to h_1 h_1$,
where $h_2$ is predominantly CP-odd.
This decay is forbidden in the CP-conserving limit
(i.e.~for $\alpha_3 = \pi/2$), but becomes allowed
when $\alpha_3 \ne \pi/2$, thereby shaping the upper
and lower bounds of the viable parameter space
in \cref{fig:smoke_inverted}.
In contrast, in the benchmark scenario \textbf{BP-SG}
discussed in \cref{sec:bp-sg}, $h_2$ was mainly CP-even,
and its decay to a pair of 125~GeV Higgs bosons is
suppressed near the alignment limit but becomes possible
for $s_{\beta - \alpha_1} \ne 0$, regardless of whether
CP is conserved. In that case, the resonant di-Higgs
searches primarily excluded parameter points on the
left and right edges of the allowed region.

As in the smoking gun scenario \textbf{BP-SG}
discussed above, in the still allowed parameter
regions in the inverted smoking gun scenario
\textbf{BP-ISG}, the current most sensitive
LHC search is the smoking gun searches for
$h_3 \to h_2 Z$ with $h_2 \to t \bar t$.
In these searches, the experimental strategies
carried out by ATLAS~\cite{ATLAS:2023zkt}
and CMS~\cite{CMS:2024yiy} do not allow
for a distinction between the CP properties of
the Higgs bosons involved.
Therefore, in the CP-conserving limit, a potential
observation in this channel could originate from
either $A \to ZH$ or $H \to ZA$ decays, assuming
both processes correspond to the same total
cross section. Recently, a proposal has been
put forward to gain sensitivity to the CP
nature of the Higgs bosons by exploiting angular
observables sensitive to the spin correlations
of the $t \bar t$ system produced in the decay of the
lighter Higgs boson~\cite{Arco:2025ydq}.
Our results strengthen the case
for the scenario
where the heavier resonance is mainly CP-odd
and the lighter is mainly CP-even,
as it 
could explain the matter-antimatter asymmetry
and predict detectable
GW signals, whereas
the scenario with the inverted
CP assignments \textbf{BP-ISG}
is not able to.
Hence, being able to experimentally distinguish
between the two in the most sensitive channel
is of key relevance to understand the physics
underlying EW symmetry breaking.

The parameter plane \textbf{BP-ISG} with
inverted smoking gun remains viable due to a slight
excess in the ATLAS data from $h_3 \to Z h_2$ searches
around masses of 450~GeV
and 650~GeV for the two spin-0 resonances, which
weakens the cross section limits~\cite{ATLAS:2023zkt}.
However, the predicted cross sections are
approximately equal to the 95\% confidence level
limits recently reported by CMS~\cite{CMS:2024yiy},
which are still not finally published and therefore
not yet included in \texttt{HiggsTools}.
Moreover, the direct production of $h_2$
decaying via $h_2 \to t \bar t$ contributes
to $t \bar t$ and $t \bar t t \bar t$ final states.
The investigated benchmark plane is compatible
with the first-year Run~2 results from CMS
$t \bar t$ searches~\cite{CMS:2019pzc}, but
the CP-conserving limit $\alpha_3 \approx \pi / 2$
would be in tension
with the full Run~2 ATLAS~\cite{ATLAS:2024vxm}
and CMS results~\cite{CMS-PAS-HIG-22-013}.
However, these updated limits are not included here
since they can only be applied under the assumption
of CP conservation where $h_2$ and $h_3$ do not interfere.
As shown in Ref.~\cite{Bahl:2025you},
interference effects between the two scalars can be
significant in the presence of CP violation and must
be properly taken into account.
Also there are significant uncertainties in modeling the
SM $t \bar t$ background near the di-top threshold.
In particular, CMS presented limits both with and
without a toponium bound state contribution, leading
to fundamentally different exclusions.
Additionally, the ATLAS and CMS results show
substantial discrepancies that remain unresolved.
Theoretically more robust limits are available from searches
in the $t \bar t t \bar t$ final state using
the full Run~2 data carried out by both
ATLAS~\cite{ATLAS:2024jja} and CMS~\cite{CMS:2019rvj},
excluding values of
$t_\beta \lesssim 1.2$ for a pseudoscalar
Higgs boson at a mass of about 450~GeV.
A similar lower limit is obtained from searches
for the charged scalars $H^\pm$ via the
decay $H^\pm \to t b$~\cite{ATLAS:2021upq}.
Considering that the benchmark plane
\textbf{BP-ISG} barely
escapes the limits from several different
LHC searches, we expect
that the inverted smoking gun scenarios
will be almost fully probed by Run~3 LHC
searches. The only way to avoid these searches
would be to increase $t_\beta$,
while at the same time we would have to decrease the mass splitting
between $h_2$ and $h_3$ to satisfy perturbative
unitarity constraints.
However, this
would further suppress the strength of
the EWPTs
and the amount of the predicted BAU,
which in the studied benchmark plane
is already below the observed value.

To summarise, although
the inverted scenario \textbf{BP-ISG}
still supports a strong EWPT but it is much less
favourable for EWBG and does not predict detectable
GW signals.
The combination of weaker phase transitions,
suppressed BAU generation, and tighter LHC (and
EDM constraints) suggests that achieving viable EWBG in
this regime would require
new dynamics to enhance the strength of the EWPT
and/or additional CP-violating sources not
contained in the ($\mathbb{Z}_2$ symmetric) C2HDM.

\subsubsection{Aligned smoking gun scenario}
\label{sec:bp-asg}

In this section, we investigate a third benchmark plane,
which we refer to as the aligned smoking gun scenario
\textbf{BP-ASG}. Unlike the previous two benchmark
planes, where the mixing angle $\alpha_1$ was varied,
here we fix $\alpha_1$ by imposing the alignment condition
$s_{\beta - \alpha_1} = 0$, which ensures SM-like
couplings of the 125~GeV Higgs boson in the
CP-conserving limit $\alpha_3 \approx 0$.
We maintain a sizable mass gap between
the two heavier neutral scalars $h_2$ and $h_3$
to facilitate a strong EWPT, as it was the case in the
original smoking gun scenario \textbf{BP-SG}
discussed in \cref{sec:bp-sg}.
Instead of varying $\alpha_1$, we explore different
values of $t_\beta$, motivated by the interplay
between collider constraints and EWBG.
While small $t_\beta$ is favoured for successful
EWBG due to enhanced top-quark transport,
it also leads to larger production cross sections
of the BSM scalars at the LHC. To study the impact of
CP violation in this context, we additionally vary the
CP-violating angle $\alpha_3$, as in the
two benchmark planes discussed above.

\begin{figure*}
\subfloat[
  \textbf{BP-ASG}: aligned smoking-gun scenario for
  a strong first-order EWPT with a mass gap between
  a lighter and mainly CP-even state~$h_2$
  and a heavier mainly CP-odd state~$h_3$.]
  {
    \label{fig:smoke_aligned}
    \includegraphics[width=0.62\textwidth]{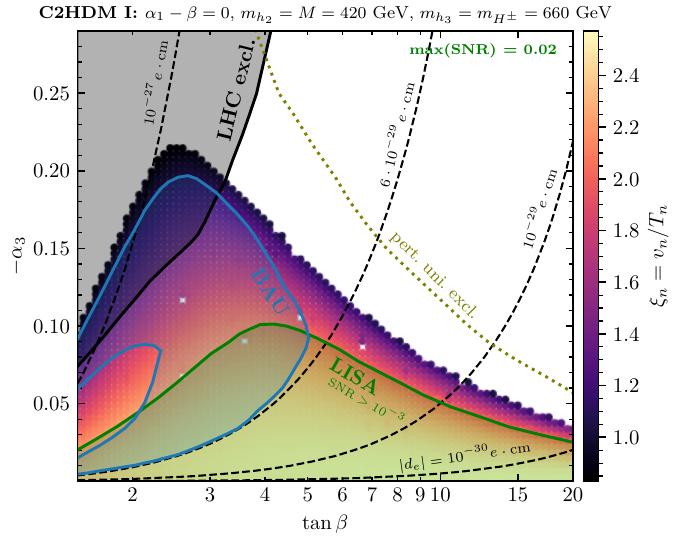}
  } \\[1em]
\subfloat[
  \textbf{BP-AMD}: scenario with algmost mass
  degenerate BSM states and a strong first-order
  EWPT]
  {
    \label{fig:almost_massdeg}
    \includegraphics[width=0.62\textwidth]{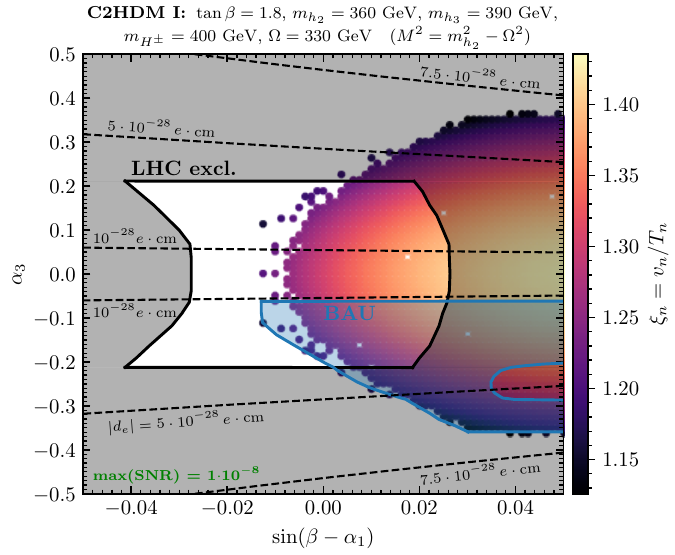}
}
\caption{
Shown are parameter points with a strong
first-order EWPT in a $\{ t_\beta ,\alpha_3\}$ plane
in the upper plot and  a $\{ s_{\beta - \alpha_1},
\alpha_3\}$ plane in the lower plot.
The colour coding of the points and the grey, blue
and green regions are defined as in \cref{fig:cpsmoker1}.
In the lower plot no green
region is visible because the EWPT is not sufficiently
strong to give rise to a detectable GW signal.}
\label{fig:cpsmoker2}
\end{figure*}

We define
the benchmark plane \textbf{BP-ASG} as
\begin{align}
\textrm{\textbf{BP-ASG:} }&1.5 \leq t_\beta \leq 20 \, ,
\; -0.3 \leq \alpha_3 \leq 0.0 \, , \notag \\
&s_{\beta - \alpha_1} = 0, \; m_{h_{2}} = M = 420\GeV \, , \notag \\
&m_{h_{3}} = m_{H^\pm} = 660\GeV \, .
\end{align}
The results for this benchmark plane are
shown in \cref{fig:smoke_aligned}.
As in the original smoking gun scenario
\textbf{BP-SG} discussed in \cref{sec:bp-sg},
the strongest EWPTs are found in the CP-conserving
alignment limit, i.e.~at $\alpha_3 = 0$,
across the entire $t_\beta$ range.
In this limit, the strength of the EWPT is independent
of $t_\beta$ due to the condition $m_{h_2} = M$.
For fixed $t_\beta$, increasing $|\alpha_3|$ leads to
a gradual weakening of the phase transition,
in line with the behaviour observed in the
benchmark plane \textbf{BP-SG}, see \cref{fig:smoke_normal}.
We find that for $t_\beta \approx 2.5$ the presence
of a strong EWPT is compatible with the largest
amount of CP violation in the Higgs sector:
strong transitions are still possible for values of
$|\alpha_3| \approx 0.2$ (with $\alpha_3 < 0$
to ensure the correct sign of the BAU).
For smaller $t_\beta \approx 1.5$, the possible
range shrinks to roughly $|\alpha_3| < 0.1$,
and for values of $t_\beta \approx 20$ it is further
reduced to approximately $|\alpha_3| \lesssim 0.03$.
All points predicting a strong EWPT in this
benchmark plane are found to be consistent with
theoretical constraints from tree-level perturbative unitarity,
which excludes only the upper right corner as indicated
with the dotted olive line.

The blue shaded region in \cref{fig:smoke_aligned}
indicates the parameter space where the predicted
BAU lies within a factor of two of the observed value,
which we take as the acceptable range given
theoretical uncertainties; see the discussion in
\cref{sec:ewbg}. One can see that successful EWBG
is possible for $t_\beta$ values roughly
up to values of $t_\beta \approx 5$.
At low $t_\beta \lesssim 2.3$,
there is a hole in
the blue region
where the predicted BAU
exceeds the upper bound of our assumed uncertainty band,
i.e.~it is more than twice the observed value.
For larger values of $t_\beta \gtrsim 5$,
on the other hand, the top transport becomes too
suppressed to generate the observed asymmetry,
and the predicted BAU falls below the acceptable range.
The presence of this upper limit on $t_\beta$ for
successful EWBG defines a clear target region for
LHC searches probing the extended Higgs sector
of the C2HDM, as will be discussed in detail below.

As with other benchmarks, the electron EDM poses a
strong constraint. The dashed black contours
indicate values exceeding $|d_e| = 6 \cdot
10^{-29}e\cdot\text{cm}$ throughout the parameter region
suitable for EWBG. This again implies that,
in the absence of either a cancellations or other
suppression mechanisms, the C2HDM
parameter space region compatible with EWBG
would be ruled out by the recent experimental
upper limits on the electron EDM.

The prospects for detecting a GW signal at LISA are
illustrated by the green-shaded region in
\cref{fig:smoke_aligned}, which corresponds to parameter
points where the EWPT generates a GW signal with a
SNR above 0.001. As discussed in \cref{sec:results},
this threshold is chosen to account for
the significant theoretical uncertainties that
affect GW predictions from cosmological phase
transitions~\cite{Croon:2020cgk}. The green region spans the entire
$t_\beta$-range shown in the plot.
As expected, the largest SNRs occur in the CP-conserving
limit at $\alpha_3 \approx 0$, corresponding to
the alignment limit, $s_{\beta - \alpha_1} = 0$,
where the EWPTs are the strongest.
For non-zero values of $\alpha_3$, the detectable GW
region extends at most up to $|\alpha_3| \approx 0.1$
at $t_\beta \approx 4$. For $t_\beta \lesssim 5$,
there is an overlap between the blue (BAU-favoured)
and green (GW-detectable) regions,
indicating that in this part of parameter space the
EWPT could simultaneously account for the observed
BAU and produce a GW signal potentially within reach
of LISA. As discussed further below, this overlapping
region is also expected to be testable at the LHC
via searches for additional Higgs bosons,
offering a compelling interplay between the LHC
and LISA in probing the nature of the EWPT and
EWBG.
However, this region is in significant tension with
the most recent experimental upper limits on the
electron EDM, as discussed above, which should be
taken into account when interpreting these results.

For larger $t_\beta$-values, the green region extends
beyond the blue one, especially for small CP-violating
phases with $\alpha_3 \approx 0$, reflecting the fact
that the strongest transitions occur in the CP-conserving
limit, where the phase transition dynamics are largely
independent of $t_\beta$. Consequently, GW signals
can remain detectable even when the BAU cannot be
generated. In the context of the type~I
Yukawa structure considered here, probing these large
$t_\beta$-scenarios at the LHC becomes increasingly 
challenging because one approaches the so-called
fermiophobic limit, where the
main production cross sections of the
BSM Higgs bosons are strongly suppressed.
In contrast, other Yukawa types feature enhanced
couplings to bottom quarks and/or tau leptons at large
$t_\beta$, potentially allowing the LHC to probe
this parameter space region via the decays of the
heavy Higgs bosons into $b \bar b$ and $\tau^+ \tau^-$
final states.
Therefore, especially in the C2HDM type~I,
it is well possible
that a GW signal compatible with an EWPT in the
C2HDM could be detected by LISA without any
possibility for a complementary discovery of
a BSM Higgs boson at the LHC.
The only collider trace of the EWPT would then
be an enhanced self coupling of the 125~GeV Higgs boson
that might be accessible at the high-luminosity
phase of the LHC, see the discussion in \cref{sec:lhc}.

We close the discussion of the benchmark plane
\textbf{BP-ASG} by
addressing the key question of how well the BAU-favoured
region can be tested at the LHC, both now and in future 
runs. The currently excluded parameter space is indicated
by the grey shaded region in the upper left corner of the
plot, which overlaps with the region favoured for EWBG
at $t_\beta \lesssim 3$ and $|\alpha_3| \gtrsim 0.06$.
This exclusion originates from ATLAS searches for
the decay of the state $h_2$ into
a $Z$ boson and a 125~GeV Higgs boson~\cite{ATLAS:2022enb}.
These searches lose sensitivity at smaller values
of $|\alpha_3|$, since the tree-level
coupling $h_2 h_1 Z$ vanishes in the alignment limit,
and they become less effective at larger $t_\beta$
due to the approximate $1/t^2_\beta$ suppression of
the gluon-fusion production cross section of $h_2$.
In most parts of the remaining viable region that
accounts for the BAU, the currently most sensitive probe
is the smoking gun signature $h_3 \to h_2 Z$,
with $h_2$ decaying into a
top-quark pair~\cite{ATLAS:2023zkt,CMS:2024yiy}, in
agreement with the results in the smoking-gun
scenario \textbf{BP-SG} discussed in \cref{sec:bp-sg}.

Looking ahead, a central question is whether the LHC
will be able to fully cover the C2HDM parameter
space that can realise EWBG, which
we here find to span up to values of
$t_\beta \approx 5$.
Based on the projections from Ref.~\cite{CMS:2025hfp},
searches in the four-top final state are expected
to significantly improve sensitivity to the state
$h_2$ at 420~GeV. While current limits from these
searches exclude values of $t_\beta \lesssim 1$ for
a dominantly CP-even $h_2$,
the projected improvement of cross section limits by
a factor of about 4-5 at the high-luminosity LHC
would extend coverage up to $\tan\beta \approx 2\text{--}3$.
Similar reach in $t_\beta$ can be expected from searches
in the di-top final state, though their sensitivity
depends more strongly on the CP nature
of $h_2$~\cite{Bahl:2025you}.
Searches for the charged Higgs bosons may also be
relevant in this regime,
currently excluding in this scenario parameter
space with $t_\beta \lesssim 1.2$ for a mass of
about 660~GeV~\cite{ATLAS:2021upq},
although the lack of
official projections prevents a definitive assessment.
For $t_\beta \lesssim 3$, all these channels could compete 
in sensitivity with the smoking gun process.
At higher $t_\beta$, however, the smoking gun channel
becomes increasingly dominant, as the
slight increase of the branching ratio
for $h_3 \to h_2 Z$ partially compensates
the $1/t^2_\beta$ drop in $h_3$ production.
To estimate the future reach of the smoking gun search,
we rescale the Run~2 ATLAS limits~\cite{ATLAS:2023zkt}
using the expected increase in statistics by a factor
of $\sqrt{3000~\mathrm{fb}^{-1}/140~\mathrm{fb}^{-1}}$,
corresponding to the expected
integrated luminosity at the end of the
high-luminosity LHC.
This extrapolation suggests that the BAU region could
be probed up to $\tan\beta \approx 4$.
When taking into account the slightly higher
center-of-mass energy at the high-luminosity
LHC and potential systematic improvements (see also 
Ref.~\cite{Arco:2025ydq}), it appears realistic that
the smoking gun searches may ultimately probe the
entire C2HDM parameter space region suitable for EWBG
in benchmark scenarios
in which the presence of the EWPT relies on the
hierarchical spectrum with
a lighter dominantly
CP-even BSM Higgs boson and a heavier dominantly
CP-odd BSM Higgs boson.
Moreover,
we verified that if the nucleation wall
width $L_w T_n$ were significantly underestimated
in our analysis due to
neglecting the stretching from plasma friction
and pressure (see the discussion in
\cref{sec:ewbg}),
the EWBG-favoured region would shrink
only mildly. In particular, the maximal value
of $t_\beta$ compatible with successful EWBG in the scenario
\textbf{BP-ASG} would decrease from
$t_\beta \approx 5$ to $t_\beta \approx 4$.
This reduction would make the prospect of probing
the entire viable region at the LHC slightly
more promising, but leaves our conclusions
otherwise unchanged.

\subsubsection{Almost mass-degenerate scenario}
\label{sec:bp-amd}

The fourth and final benchmark plane we define
represents a qualitatively different scenario from
those analysed previously. The benchmark
planes \textbf{BP-SG}, \textbf{BP-ISG} and
\textbf{BP-ASG} relied on a sizable mass splitting between
the additional neutral Higgs bosons to induce a
strong first-order EWPT.
In contrast, this scenario features an approximately
degenerate spectrum for the neutral and
charged BSM scalars. In this setup, the potential barrier
between the EW symmetric and broken phases
required for a strong EWPT stems from
sizable quartic couplings that arise due to a
separation between the physical scalar masses and
the soft $\mathbb{Z}_2$-breaking scale $M$
defined in \cref{eq:M}.
This mechanism is motivated by the detailed study
in Ref.~\cite{Dorsch:2017nza}, where the relevant conditions
for realising a strong EWPT in the 2HDM with a
degenerate scalar spectrum were systematically explored
and contrasted with the case of hierarchical mass spectra.
Additional support for this setup comes from
Ref.~\cite{Basler:2017uxn}, where explicit C2HDM benchmark
points with approximately degenerate BSM scalars were
found that featured a strong EWPT and satisfied
Run~1 LHC and electron EDM constraints
available at the time.

Specifically, we consider a benchmark plane
denoted as the almost mass-degenerate
(\textbf{BP-AMD}) scenario defined by
\begin{align}
\textrm{\textbf{BP-AMD:} }&-0.05 \leq s_{\beta - \alpha_1} \leq 0.05 \, , \;
-0.5 \leq \alpha_3 \leq 0.5 \, , \notag \\
&t_\beta = 1.8\, , \; m_{h_{2}} = 360\GeV \, , \notag \\
&m_{h_3} = 390\GeV \, , \;
m_{H^\pm} = 400\GeV \, , \notag \\
&\Omega = m_{h_2}^2 - M^2 = 330\GeV \, ,
\end{align}
where the parameter $\Omega$
was defined in Ref.~\cite{Dorsch:2016nrg}
to parametrise the splitting between the
scale $M$ and the almost degenerate Higgs bosons masses
$m_{h_2}$, $m_{h_3}$ and~$m_{H^\pm}$.

The parameter values defining this benchmark plane
are carefully chosen to optimise the conditions for a
strong EWPT within a degenerate scalar spectrum while
remaining consistent with theoretical and LHC
constraints. As in the previously discussed
benchmark planes \textbf{BP-SG} and \textbf{BP-ISG},
we vary the mixing angles $\alpha_1$ and $\alpha_3$,
where $\alpha_1$ is scanned around the alignment limit
condition $s_{\beta - \alpha_1} = 0$.
The scan range for $\alpha_3$ is taken to be substantially
broader than in the previous benchmark planes.
This is possible because, in the present scenario,
and in contrast to hierarchical mass spectra,
the near-degeneracy of the neutral scalars $h_2$
and $h_3$ allows for sizable CP admixtures without
violating theoretical constraints.
The value of $t_\beta = 1.8$ is chosen to be as small
as feasible, since lower values generally enhance
the prospects for a strong EWPT by allowing larger
values of the splitting parameter $\Omega$
without conflicting with perturbative unitarity
constraints~\cite{Dorsch:2017nza}.
However, values of $t_\beta$ much below
the chosen value are
ruled out by LHC searches for additional Higgs bosons,
in particular from searches for the charged
scalars~\cite{ATLAS:2021upq}.
The BSM Higgs bosons are required to
remain relatively light in this scenario to ensure
sufficiently large non-decoupling effects which
are essential for generating a strong EWPT
while not being in tension with perturbative unitarity
and vacuum stability constraints~\cite{Dorsch:2017nza}.
Consequently, all BSM Higgs boson masses are chosen just above
the di-top threshold, where viable parameter space remains
because LHC searches in the
$\tau^+ \tau^-$~\cite{CMS:2024ulc,ATLAS:2020zms},
$W^+W^-$~\cite{ATLAS:2017uhp,CMS:2019bnu}
and $ZZ$~\cite{CMS:2018amk,ATLAS:2020tlo}
final states start to lose sensitivity in type~I.
Finally, after fixing the angles and masses,
the splitting parameter $\Omega$ is set to a value
inspired by the analysis in Ref.~\cite{Dorsch:2017nza},
such that a strong EWPT is obtained in the
alignment limit.
At the same time, the value of $\Omega$ is kept
moderate enough to avoid generating quartic scalar
couplings in conflict with perturbative
unitarity bounds or vacuum stability constraints.

The results for the benchmark plane \textbf{BP-AMD}
are shown in \cref{fig:almost_massdeg}.
Despite the absence of large mass hierarchies,
which 
facilitates the strongest EWPTs,
this scenario can still realise a
strong first-order EWPT.
As before, the colour coding of the points
in the $\{s_{\beta - \alpha_1}, \alpha_3\}$ plane
indicates the strength of the phase transition through
the value of $\xi_n$.
We observe that the strongest transitions $\xi_n \approx
1.4$ are reached in the CP-conserving limit
for low values of $\alpha_3 \approx 0$.
The values of $\xi_n$ that can be achieved here
are significantly smaller than the maximum
values of $\xi_n \approx 3$ that we found in
scenarios with a hierarchical spectrum, see the
scenario \textbf{BP-SG} shown in \cref{fig:smoke_normal},
but they satisfy the baryon-number preservation
condition shown in \cref{eq:bnpc}. 
This demonstrates that the EWPTs predicted in 
an almost mass degenerate scenario
might be sufficiently strong for a realising 
EW baryogenesis.

The blue contour marks the region where the
predicted baryon-to-entropy ratio $\eta_s$
is compatible with the observed value
within an uncertainty of a factor of two.
This viable BAU region arises despite the near-degeneracy of the scalar states, suggesting that successful EWBG does not strictly require a large mass gap between the
neutral BSM Higgs bosons in the C2HDM.
Accordingly, the smoking gun
signature $h_3 \to Z h_3$ is not distinctive of
this scenario (see also the discussion
in Refs.~\cite{Basler:2017uxn}).
However, as we will discuss in detail
below, 
other LHC searches 
can probe and likely exclude (or confirm) this 
benchmark plane in the near future.

The parameter space region viable for EWBG
is again in tension with the most recent constraints
on the electron EDM. The dashed black lines show
constant contours $|d_e|$, which exceed
$|d_e| \approx 10^{-28}\, e \cdot \text{cm}$ in the blue
region preferred by EWBG, well above the experimental bound
of $|d_e| < 4.1 \cdot 10^{-30}\, e\cdot\text{cm}$.
The tension with the non-observation of an
electron EDM in this scenario is about a factor
of two more severe compared to the scenarios
\textbf{BP-SG} and \textbf{BP-ASG}.
This increased tension can be attributed to the
overall lighter BSM Higgs boson spectrum
in agreement with expectation following
\cref{eq:scaling_edm}.
Moreover, in comparison to \textbf{BP-SG}, the smaller
value of~$t_\beta$ leads to larger predictions
for the electron EDMs.

The limited strength
of the EWPTs places the predicted GW signals below
the sensitivity of the LISA experiment.
The predicted GW signal is extremely suppressed.
The maximum SNR in LISA for this benchmark is
$\text{SNR}_{\max} \approx 10^{-8}$, which is much
below the sensitivity threshold of LISA and is effectively
unobservable in future space-based GW interferometers.
This aligns with expectations that compressed spectra
tend to yield weaker transitions,
smaller latent heat and lower bubble
wall velocities, reducing the efficiency of
GW production.
This implies that any hints that might be revealed
in future LHC searches consistent with this
scenario would point to an EWPT without a corresponding
stochastic GW background detectable
at LISA.

The LHC constraints on this benchmark plane are
illustrated by the grey-shaded exclusion regions
and the solid black contour enclosing the remaining
viable parameter space. The excluded regions to the
left and right of the allowed area arise
from searches for the state $h_2$, targeting
the decay $h_2 \to ZZ$~\cite{ATLAS:2020tlo}, which become
increasingly sensitive as one departs
from the alignment condition $s_{\beta - \alpha_1} = 0$.
At the mass of $h_2 = 360\GeV$ considered here,
we find these searches to be more constraining
than the searches for resonant Higgs boson pair
production via $h_2 \to h_1 h_1$ decays.
The parameter space regions above and below the
allowed region are excluded by searches targeting
the $h_2 \to h_1 Z$ decay mode~\cite{ATLAS:2022enb}.
In this benchmark
plane, these searches begin to be
the most sensitive search channel
for $h_2$ at values of $|\alpha_3| \gtrsim 0.05$.
One can see that the two searches responsible
for the shape of the
exclusion region already exclude a substantial
part of the blue region that is favoured by EWBG.

The parameter choices in the benchmark plane
\textbf{BP-AMD} were deliberately optimized to
evade existing LHC bounds, but the remaining viable
region is expected to be comprehensively tested as
the LHC collects more data. Several search channels,
while not currently excluding this scenario,
lie just below the sensitivity threshold and will
soon become decisive. In particular, searches for the
charged Higgs bosons through the $H^\pm \to t b$ decay
already constrain slightly smaller values of $t_\beta$,
and for the chosen value of $t_\beta = 1.8$ the predicted
cross sections are only about 10\% below the current
95\% confidence level limits~\cite{ATLAS:2021upq}.
As a result, charged Higgs boson searches are likely
to probe this almost mass degenerate C2HDM scenario
even at moderately larger $t_\beta$ (or $m_{H^\pm}$),
though EWBG restricts $t_\beta$ from becoming too large.
Moreover, very close to and within the alignment
limit, the channels
$h_{2,3} \to t \bar t$~\cite{CMS:2019pzc},
$h_2 \to \tau^+ \tau^-$~\cite{ATLAS:2020zms,CMS:2022goy}
and $h_3 \to \gamma\gamma$~\cite{ATLAS:2021uiz}
become the most
sensitive probes, filling in the gaps left by the
searches in the $ZZ$ and $h_1 Z$ final states
responsible for the exclusion contours,
which lose sensitivity in this region.
These channels offer the exciting prospect of
observing not only a single additional Higgs boson,
but potentially the entire extended scalar sector
predicted in such a scenario.
Additional coverage outside of the
alignment limit is also expected from searches
for resonant pair production of the SM-like Higgs boson
via $h_2/h_3 \to h_1 h_1$ decays~\cite{CMS:2018ipl,
ATLAS:2023vdy}.
Altogether, the current configuration of this
benchmark represents a carefully chosen
and narrow window that has thus far escaped exclusion
from LHC searches. However, with ongoing and future
LHC runs it appears likely that the remaining gaps
will be closed, potentially already with the
data collected during Run~3,
making the benchmark \textbf{BP-AMD} a fully
testable realisation of EWBG.

In summary, \textbf{BP-AMD} highlights that viable
EWBG remains possible even in the absence of large
scalar mass splitting, though such scenarios face
major challenges from stringent LHC constraints that
will soon confirm or exclude if such a C2HDM
scenario is realised in nature.
The fact that the LHC can probe these scenario
is crucial since the predicted GW signals are significantly
below the expected sensitivity of future
space-based GW experiments, which therefore will not
be able to probe such a scenario.

\section{Summary and conclusions}

In this work, we have presented an analysis of the
EWPT and the possibility to explain the
imbalance between matter and antimatter
within the C2HDM, a 2HDM with softly broken
$\mathbb{Z}_2$ symmetry and explicit CP violation
in the Higgs potential.
The C2HDM is the simplest scalar
extension of the~SM that features
new sources of CP violation and can accommodate
a strong first-order EWPT, both vital ingredients
for a explaining the matter-antimatter asymmetry 
through EWBG.
Rather than performing fully general parameter
scans, we have focused on defining well-motivated benchmark
planes, which enables a more transparent analysis
of how CP violation influences the dynamics of
the EWPT in the C2HDM compared to the
extensively studied (CP-conserving) R2HDM.
This also allows for a more illustrative analysis of the
impact of the key
LHC search channels that can probe the still allowed
parameter space regions relevant for successful EWBG.
The benchmark planes are intended to guide
ongoing efforts in defining representative scenarios
for the exploration of CP-violation in extended scalar
sectors at the LHC Run 3 and beyond,
while remaining consistent with current collider
constraints and theoretical requirements.

It should be noted that we have not imposed the
electron EDM bounds as a hard constraint.
Taking into account the recent experimental
upper bounds, this
would leave barely any room for realising EWBG
in the C2HDM.
Instead, we evaluated and displayed the predicted
electron EDM values across the benchmark planes in
order to quantify the degree of tension with
the most recent experimental limits.
This reflects the possibility that the electron EDM
may be suppressed in more general extensions of
the C2HDM, without altering the dynamics of
the EWPT.

Primordial GW signals offer an additional,
complementary probe of the EWPT.
We have assessed the potential detectability
of a stochastic GW background produced from an EWPT
within the C2HDM, using the predicted SNR
for the LISA experiment as a reference. 
Taking into account the sizeable theoretical uncertainties
associated with the modelling of the phase transition
dynamics and GW production, we used a lower SNR
threshold of 0.001 instead of the more conventional
value of 1. This choice reflects the challenges in
accurately predicting the strength of GW signals from
the EWPT, while still allowing us to identify regions
of parameter space where detectable signals could
potentially be observed by LISA.
One of the key questions addressed in our work
is whether EWBG and detectable GW signals can be
simultaneously realised, and if so, which parameter space
regions can accommodate both. Furthermore, we examined
the LHC signals that correspond to these regions
in order to explore the potential interplay between
the LHC and space-based GW astronomy in probing
EWBG in the next decade.

In total, we explored four distinct benchmark
planes: the smoking gun scenario \textbf{BG-SG}
shown in \cref{fig:smoke_normal},
the inverted smoking gun scenario \textbf{BP-ISG}
shown in \cref{fig:smoke_inverted},
the aligned smoking gun scenario \textbf{BP-ASG}
shown in \cref{fig:smoke_aligned},
and the almost mass degenerate scenario \textbf{BP-AMD}
shown in \cref{fig:almost_massdeg}.
The first three benchmark scenarios share the feature
that the presence of the strong EWPT relies on
a sizeable mass splitting between the neutral BSM
Higgs bosons $h_2$ and $h_3$ contained in the C2HDM.
In the fourth
benchmark scenario all
neutral and charged BSM
Higgs bosons $h_2$, $h_3$ and $H^\pm$ are chosen
to be close in mass,
and the EWPT is facilitated by a splitting between
the $\mathbb{Z}_2$-breaking scale~$M$ and the
physical masses of the Higgs bosons.

Among the four benchmark planes,
all but the inverted smoking
gun scenario \textbf{BP-ISG} feature
parameter regions in which
the observed baryon asymmetry can be explained
by EWBG. 
The parameter space regions favoured by
EWBG show a tension
with the recent experimental upper bounds on
the electron EDM by about an order of magnitude
or more.
This tension is most severe in the almost mass
degenerate scenario \textbf{BP-AMD} due to an
overall lighter Higgs spectrum.

Regarding GWs,
only the smoking gun scenario \textbf{BP-SG}
and the aligned smoking gun scenario \textbf{BP-ASG}
yield SNRs at LISA of the order of
$0.001\text{--}10$, making
them the only scenarios that can potentially be probed with
space-based GW experiments.
This is consistent with and confirms earlier
findings in the R2HDM, where it was shown that a mass
hierarchy featuring a lighter BSM CP-even scalar $H$
and a heavier CP-odd scalar $A$ leads to the
strongest EWPTs, motivating the designation of the
associated $A \to ZH$ decay channel as
a smoking gun signature at the LHC.
Within these two benchmark planes, we have found that
parts of the region favoured by EWBG and of the
region predicting potentially detectable
GW signals overlap,
promising an interesting interplay between the LHC
and LISA in probing EWBG in the C2HDM.

We have shown that, in the presence of CP violation,
the strength of the EWPT tends to be reduced
compared to the CP-conserving case. 
Nevertheless, there remains sufficient parameter space
with significant amount of
CP violation in the Higgs sector
that satisfy the baryon number preservation condition and
predicts potentially detectable GW signals.
Importantly, in the \textbf{BP-SG} scenario
we have shown that introducing CP-violating
mixing (while keeping other parameters fixed)
can help avoid vacuum trapping in the EW-symmetric
phase, which would otherwise prevent a phase
transition from occurring. As a result, the
impact of this vacuum trapping effect, which
was observed to frequently prevent the potentially
strongest transitions in the R2HDM,
can be mitigated in the C2HDM.

On the contrary to the two (aligned) smoking gun
scenarios \textbf{BP-SG} and \textbf{BP-ASG},
for the almost mass degenerate
scenario \textbf{BP-AMD} and the inverted
smoking gun scenario \textbf{BP-ISG}
we have found that the
predicted SNRs remain below the nominal sensitivity
of LISA by 8 and 10 orders of magnitude, respectively.
This is a consequence of the limited strength of
the EWPT in these scenarios. Nevertheless, for
the almost mass degenerate scenario \textbf{BP-AMD},
we have shown that one can accommodate an EWPT that
is sufficiently
strong to facilitate EWBG, although this possibility
is already under strong pressure from the non-observation
of additional Higgs bosons at the LHC.
Since the GW signals predicted in this scenario
are far from reaching the detectability threshold
of LISA, our results emphasize 
that observable GWs are not a generic prediction of
EWBG in the C2HDM. Due to the tiny SNRs of
the order of $10^{-8}$ or below, this conclusion is
unlikely to be affected
by the large uncertainties related to the description
of cosmological phase transitions and the
resulting stochastic GW background.
In such cases, the role of the LHC in probing
the nature of EW symmetry breaking is especially
critical.

Across all four benchmark scenarios considered in
this work, current LHC searches already constrain
significant portions of the parameter space compatible
with a strong first-order EWPT. In the smoking gun
scenario \textbf{BP-SG}, wide regions are excluded
already by searches for resonant Higgs boson
pair production ($h_2 \to h_1 h_1$) and
the production of a Higgs boson in association with
a $Z$ boson ($h_2 \to Z h_1$),
with the smoking gun channel $h_3 \to h_2 Z$
(with $h_2 \to t\bar{t}$) remaining the most
sensitive probe in the viable region,
centered around the alignment limit.
A similar pattern holds in the inverted scenario
\textbf{BP-ISG}, which only narrowly escapes current
limits due to a small excess in ATLAS data
from the smoking gun searches in the $Z t \bar{t}$ final state.
The inverted scenario requires smaller
$t_\beta$ values to facilitate a strong EWPT,
rendering it well
within reach of several LHC searches,
and is likely to be fully tested during Run 3.
The aligned smoking gun scenario \textbf{BP-ASG}
shows that successful EWBG is possible up to
$\tan\beta \approx 5$, with the smoking gun and
multi-top signatures, in combination with
searches for the
production of the 125~GeV Higgs boson from
heavier spin-0 resonances, again dominating current
and projected sensitivity.
Meanwhile, the almost mass-degenerate scenario
\textbf{BP-AMD}, which was carefully tailored to
evade current LHC bounds, lies within reach of
various LHC searches targeting the decays
$h_{2,3} \to h_1 h_1$,
$h_{2,3} \to ZZ$, $h_{2,3} \to Z h_1$,
$h_{2,3} \to \tau^+ \tau^-$,
$h_{2,3} \to \gamma\gamma$, and
(most notably) $H^\pm \to tb$.

In all of the four benchmark planes,
searches involving multi-Higgs
signatures, where a heavy new Higgs boson decays
into final states including the SM-like Higgs boson
or another lighter BSM Higgs boson,
already exclude large portions of the
otherwise viable parameter space or represent the
most promising discovery channels.
This underscores their
growing importance for testing the EWPT at
current and future LHC runs. 
Crucially, it is only during Run 2, with the LHC
operating for the first time at 13~TeV center-of-mass
energy, that these multi-Higgs and cascade decay
signatures have become experimentally accessible,
opening uncharted territory in the hunt for
additional Higgs bosons.
Our results show that the crucial
overlap region, where successful EWBG and
detectable GW signals coexist, is within the
reach of the LHC, particularly through the smoking
gun decay channels. This potential interplay
between the LHC and LISA strengthens the case
for a coordinated effort to probe the fundamental origin
of electroweak symmetry breaking and the BAU.
It should be noted, however
(as illustrated, e.g.~in the aligned smoking
gun scenario \textbf{BP-ASG}), that the LHC will
not be able to cover all parameter regions predicting
GW signals detectable at LISA via searches
for additional Higgs boson,
especially (in type~I) at larger $t_\beta$.
However, larger values of $t_\beta$ suppress
the generation of the BAU.
In this case, measurements of
the Higgs boson self coupling from non-resonant
pair production remain as a final tool to
indirectly probe the parameter space during
the high-luminosity phase of the LHC.

Altogether, our findings paint a realistic picture of the current status of EWBG in the C2HDM. The parameter space that supports baryogenesis is under increasing pressure from EDM bounds and offers limited prospects for GW detection at LISA. Nonetheless, EWBG in the C2HDM remains phenomenologically interesting,
still providing
a well-motivated framework 
for probing
the nature of the EWPT during the upcoming LHC Runs.
Our benchmark planes are intended to serve as a roadmap for future experimental exploration, while reflecting the evolving experimental constraints that challenge the original motivation for EWBG in the minimal C2HDM. Whether this window can be preserved, or points instead toward extensions beyond the minimal setup, remains an open and timely question.

\section*{Acknowledgments}
We thank Howard Haber for useful discussions and
Lisa Biermann for helpful correspondence about
\texttt{BSMPT}.
TB acknowledges the support of the Spanish Agencia
Estatal de Investigaci\'on through the grant
``IFT Centro de Excelencia Severo Ochoa CEX2020-001007-S''.
The project that gave rise to these
results received the support of a
fellowship from the ``la Caixa''
Foundation (ID 100010434). The
fellowship code is LCF/BQ/PI24/12040018.
MOOR is supported by the STFC under grant ST/X000753/1.

\appendix

\section{Parameter relations}
\label{app:pararelations}

In order to obtain the Lagrangian parameters
from the set of input parameters shown in
\cref{eq:input_parameters_c2hdm},
we solve numerically for the parameters
\begin{equation}
\alpha_1,\quad \lambda_1,\quad \lambda_2,\quad \lambda_3,\quad \lambda_4,\quad \text{Re}\,\lambda_5,\quad \text{Im}\,\lambda_5,
\end{equation}
using the following system of equations, where
$s_\beta = \sin\beta$, $c_\beta = \cos\beta$ and $R_{ij}$ are the rotation matrix elements defined in \cref{eq:rmatrix},
\begin{align}
&m_{h_1}^2 R_{11}^2 + m_{h_2}^2 R_{21}^2 + m_{h_3}^2 R_{31}^2 = v^2 c_\beta^2 \lambda_1 + m_{12}^2 \tan\beta \,, \\
&m_{h_1}^2 R_{11} R_{12} + m_{h_2}^2 R_{21} R_{22} + m_{h_3}^2 R_{31} R_{32} \notag \\ &= -m_{12}^2 + v^2 c_\beta s_\beta (\lambda_3 + \lambda_4 + \text{Re}\,\lambda_5) \,, \\
&m_{h_1}^2 R_{11} R_{13} + m_{h_2}^2 R_{21} R_{23} + m_{h_3}^2 R_{31} R_{33} \notag \\ &= -\frac{1}{2} v^2 s_\beta\, \text{Im}\,\lambda_5 \,, \\
&m_{h_1}^2 R_{12}^2 + m_{h_2}^2 R_{22}^2 + m_{h_3}^2 R_{32}^2 = m_{12}^2 \cot\beta + v^2 s_\beta^2 \lambda_2 \,, \\
&m_{h_1}^2 R_{12} R_{13} + m_{h_2}^2 R_{22} R_{23} + m_{h_3}^2 R_{32} R_{33} \notag \\ &= -\frac{1}{2} v^2 c_\beta\, \text{Im}\,\lambda_5 \,, \\
&m_{h_1}^2 R_{13}^2 + m_{h_2}^2 R_{23}^2 + m_{h_3}^2 R_{33}^2 = -v^2 \text{Re}\,\lambda_5 + \frac{m_{12}^2}{s_\beta c_\beta} \,, \\
&m_{H^\pm}^2 = -\frac{1}{2} v^2 (\lambda_4 + \text{Re}\,\lambda_5) + \frac{m_{12}^2}{s_\beta c_\beta} \,.
\end{align}
The parameters $m_{11}^2$ and $m_{22}^2$ can then
be obtained via
\begin{align}
m_{11}^2 &= \frac{m_{12}^2\, v_2}{v_1} - \frac{1}{2} v_1^2 \lambda_1 - \frac{1}{2} v_2^2 \lambda_3 - \frac{1}{2} v_2^2 \lambda_4 - \frac{1}{2} v_2^2\, \text{Re}\,\lambda_5\, , \\
m_{22}^2 &= \frac{m_{12}^2\, v_1}{v_2} - \frac{1}{2} v_2^2 \lambda_2 - \frac{1}{2} v_1^2 \lambda_3 - \frac{1}{2} v_1^2 \lambda_4 - \frac{1}{2} v_1^2\, \text{Re}\,\lambda_5\, ,
\end{align}
and the imaginary part of $m_{12}^2$ depends
on $\textrm{Im}\lambda_5$ and is given
via \cref{eq:imm12sq}.
Solving the system of equations given above
for $\alpha_3 \neq 0$ leads to two physically
distinct solution, from which only one has
$\mathrm{Im}\lambda_5 \neq 0$, thus corresponding
to a CP-violating parameter point. We always pick
this CP-violating solution in our numerical
analysis.

\section{Bubble width in the C2HDM}
\label{app:Lw}

\begin{figure}[t]
  \centering
  \includegraphics[width=0.98\linewidth]{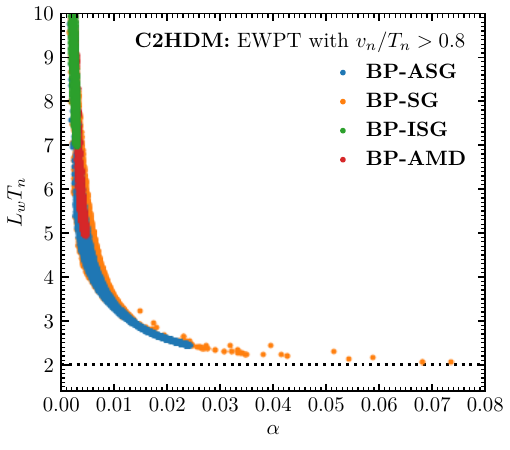}
  \caption{Predictions for $L_w T_n$ against $\alpha$
  for the benchmark planes defined in
  \cref{sec:results}. BP-ASG: aligned smoking gun scenario
  (blue), BP-SG: smoking gun scenario (orange),
  BP-ISG: inverted smoking gun scenario (green),
  BP-AMD: almost mass degenerate scenario (red).
  The horizontal dotted line indicates the
  value $L_w T_n = 2$, where the gradient expansion
  of the bubble wall profile becomes unreliable.}
  \label{fig:lwtn}
\end{figure}

We make use of the results of
Refs.~\cite{Fromme:2006wx,Dorsch:2016nrg} to
estimate the baryon-to-entropy ratio~$\eta_s$
according to \cref{eq:eta_estimate}.
This approximation for the BAU
relies on a gradient expansion
of the bubble profile,
whose validity breaks down for too thin
bubbles with $L_w T_n \lesssim 2$.
In order to verify whether the parameter points
shown in our benchmark planes satisfy this condition,
we show in \cref{fig:lwtn} the values of $L_w T_n$
that we obtain using the approach discussed in
\cref{sec:ewbg}.
One can see that the parameter points from the
benchmark planes BP-ISG and BP-AMD discussed
in \cref{sec:bp-isg} and \cref{sec:bp-amd},
respectively, feature values of $L_w T_n$
that are substantially larger then two.
The benchmark planes BP-SG and BP-ASG discussed
in \cref{sec:bp-sg} and \cref{sec:bp-asg}, respectively,
feature parameter points with stronger EWPT.
For the parameter points with the strongest transitions,
the values of $L_w T_n$ approach the limit of two,
where the application of the gradient expansion
becomes problematic.
However, this only affects a very small fraction
of the parameter points, while the majority of
points features values of $L_w T_n \gtrsim 3$ .
We therefore consider the WKB framework based on the
gradient expansion to be applicable for the bulk
of parameter space investigated here.

\bibliography{references}
\bibliographystyle{apsrev4-2}

\end{document}